\documentclass[aps,prd,superscriptaddress,nofootinbib,amsmath,amsfonts,preprintnumbers,groupedaddress,showpacs,10pt,english]{revtex4}
\usepackage{amsmath}
\usepackage{amssymb}
\usepackage{babel}
\usepackage{wrapfig}
\usepackage{cancel}

\usepackage{relsize,exscale}
\makeatletter

\usepackage{array,multirow,graphicx}
\usepackage{dcolumn}
\usepackage{newlfont}
\usepackage{bm}
\usepackage[colorlinks,citecolor=blue,urlcolor=blue,linkcolor=blue]{hyperref}
\usepackage[figtopcap]{subfigure}
\usepackage{color}
\usepackage{verbatim}

\usepackage{scalerel}
\usepackage{tikz}
\usetikzlibrary{svg.path}
\definecolor{orcidlogocol}{HTML}{A6CE39}
\tikzset{
  orcidlogo/.pic={
    \fill[orcidlogocol] svg{M256,128c0,70.7-57.3,128-128,128C57.3,256,0,198.7,0,128C0,57.3,57.3,0,128,0C198.7,0,256,57.3,256,128z};
    \fill[white] svg{M86.3,186.2H70.9V79.1h15.4v48.4V186.2z}
                 svg{M108.9,79.1h41.6c39.6,0,57,28.3,57,53.6c0,27.5-21.5,53.6-56.8,53.6h-41.8V79.1z M124.3,172.4h24.5c34.9,0,42.9-26.5,42.9-39.7c0-21.5-13.7-39.7-43.7-39.7h-23.7V172.4z}
                 svg{M88.7,56.8c0,5.5-4.5,10.1-10.1,10.1c-5.6,0-10.1-4.6-10.1-10.1c0-5.6,4.5-10.1,10.1-10.1C84.2,46.7,88.7,51.3,88.7,56.8z};}}
\newcommand\orcid[1]{\href{https://orcid.org/#1}{\mbox{\scalerel*{
\begin{tikzpicture}[yscale=-1,transform shape]
\pic{orcidlogo};
\end{tikzpicture}
}{|}}}}
\graphicspath{{./}{Figs/}}
\begin{document}
\date{\today}
\title{ Extension of Hayward black hole in   $f(R)$ gravity coupled with a scalar field }
\author{G.~G.~L.~Nashed~\orcid{0000-0001-5544-1119}}
\email{nashed@bue.edu.eg}
\affiliation {Centre for Theoretical Physics, The British University in Egypt, P.O. Box
43, El Sherouk City, Cairo 11837, Egypt\\
Center for Space Research, North-West University, Potchefstroom 2520, South Africa}
\begin{abstract}
This study looks into regular solutions in a theory of gravity called $f(R)$ gravity, which also involves a scalar field.
The $f(R)$ theory changes Einstein's ideas by adding a new function related to something called the Ricci scalar. This lets us tweak the equations that describe how gravity works.  Adding a scalar field makes the theory more interesting, giving us more ways to investigate and understand it. { The main goal of this research is to create regular black holes using a combination of $f(R)$ gravitational theory and a scalar field.}  Regular solutions don't have any singularities, which are points where certain physical quantities, like invariants, become big or undefined. { In this context, we find two regular black hole solutions by using a spherical space with either an equal or unequal approach.} For the solutions where we use the equal approach, we figure out the shape of $f(R)$ and how it changes, along with its first and second derivatives. We demonstrate that Hayward's solution in this theory stays steady because all the shapes of $f(R)$ and their first and second derivatives are positive. Next, we focus on the case where the metric isn't equal and figure out the black hole solution. We also find out what $f(R)$ and the scalar field look like in this situation. We demonstrate that the solution in this case is a broader version of the Hayward solution. When certain conditions are met, we end up back at the scenario where the metrics are equal.  We also prove that this model is stable because $f(R)$, along with its first and second derivatives, are all positive. { We analyze the trajectories of these black hole solutions and determine the forms of their conserved quantities that remain the same along those trajectories.}
\keywords{ $f(R)$ with scalar field; regular  black holes; stability; geodesics; thermodynamics.}
\pacs{ 04.50.Kd, 98.80.-k, 04.80.Cc, 95.10.Ce, 96.30.-t}
\end{abstract}
\maketitle
\section{Introduction}\label{S1}
Black holes are incredibly unusual areas in the fabric of space and time. Their key feature is the event horizon, a one-way boundary in space and time that even light can't get out of.
Black holes are known as the final result of matter collapsing under gravity and are a key prediction of Einstein's general theory of relativity \cite{Einstein:1916vd,Schwarzschild:1916uq,Penrose:1964wq}.
Furthermore, black holes could hold the key to finally combining GR and quantum mechanics, a goal scientists have been aiming for \cite{Hawking:1976ra, Giddings:2017jts} (see also \cite{Giddings:2019jwy}).
Certainly, delving deeper into the study of black holes will lead to a better understanding of gravity, especially at energy levels that we can't achieve here on Earth.

 In terms of what we can see, black holes show up in many different space situations. There's plenty of evidence, both direct and indirect, showing that supermassive black holes exist, some weighing up to $10^{10}$ times the mass of our Sun.  It's thought that supermassive black holes are usually found at the centers of most big galaxies, including our own~\cite{Lynden-Bell:1969gsv, Kormendy:1995er}. Supermassive black holes are seen as major sources of energy, outshining other parts of galaxies. They're known as active galactic nuclei because they make the center of the galaxy extremely bright \cite{Bambi:2019xzp}.

Gravitational collapse happens when massive objects like stars suddenly collapse because of gravity, creating compact objects like black holes or neutron stars. Knowing how gravity makes things collapse and form singularities is key to understanding how matter and space behave in really extreme situations \cite{Penrose:1964wq, Hawking:1970zqf, Senovilla:1998oua}.  The idea of singularities appears in different areas like math, physics, and astrophysics. They're linked to the most extreme situations in space and time. One well-known example is the singularity in the framework of GR. According to the theory, when matter collapses enough because of gravity, it creates a singularity with infinite density and curvature at the center of the black hole. This singularity is often linked to our current physics breaking down because the laws of physics we know don't work there anymore \cite{Penrose:1969pc, Wald:1997wa}.

Since Bardeen's initial research \cite{bardeen1968non}, there's been a big push to find black hole solutions that don't have singularities. To achieve this, researchers can either change how gravity works or search for unusual types of matter that can smooth out the central singularity.
For a complete list of studies in this field, check out:~\cite{Borde:1996df,Ayon-Beato:1998hmi,Burgess:2005sb,Nashed:2001cp,Hayward:2005gi,Nashed:2020kdb,Iso:2006ut,Berej:2006cc,Li:2008fa,Nashed:2018piz,DeFelice:2009rw,Bronnikov:2012ch,Rinaldi:2012vy,Bambi:2013ufa,Shirafuji:1996im,
Toshmatov:2014nya,Johannsen:2013szh,Nashed:2018qag,Sebastiani:2016ras, Nashed:2020mnp,Colleaux:2017ibe,Chinaglia:2018gvf,ElHanafy:2014efn,Han:2019lfs,Nashed:2001im,Rodrigues:2019xrc,Jusufi:2019caq,Gorji:2020ten}.
To explore important studies on how we can observe these black holes, look at~\cite{Schee:2015nua, Stuchlik:2014qja, Schee:2016mjd, Stuchlik:2019uvf, Schee:2019gki}.

Regular black holes (RBHs) are a type of black hole that have coordinate singularities, also called horizons, but they don't have the important singularities that exist throughout the entire spacetime. Usually, finding a regular black hole involves studying a spacetime where the curvature measures stay small everywhere, particularly at the center of the black hole \cite{Dymnikova:1992ux,Ayon-Beato:1998hmi,Bronnikov:2000vy}. This idea is related to Markov's proposal about limiting curvature, which suggests that curvature measures should always be restricted to a particular universal value. However, this method faces challenges when used with the well-known Taub-NUT black hole because the paths of light and matter show gaps at the horizon. This disagreement goes against the other way of creating a smooth spacetime by making sure all paths are complete\footnote{Following this idea, spacetime is seen as regular if its paths for light and matter go on indefinitely without stopping at any fixed point.}
\cite{Wald:1984rg}. Using complete paths for matter and light also has its difficulties \cite{Carballo-Rubio:2021wjq}, as shown by examples found in other studies \cite{GEROCH1968526,Olmo:2015bya}. In these cases, the paths for matter and light are complete, but the measures of curvature become infinite, which goes against Markov's idea of restricting curvature. In this situation, both methods must cooperate smoothly to accurately assess regular black holes.

Similar to regular black hole solutions, some solutions to the equations describing gravity in GR show issues like divergences or singularities. These singularities might come from limitations within the classical theory and could be resolved by looking at them from a quantum perspective. In classical physics, James Bardeen made a significant breakthrough in 1968 \cite{Bambi:2013ufa,} by discovering the first regular black hole solution in three spatial dimensions plus time. This solution avoids problems with measurements of the shape of space and how much it's curved that usually go wild. Later on, Ay$\acute{o}$n-Beato and Garc$\acute{i}$a \cite{Ayon-Beato:2000mjt} proposed an interesting idea suggesting that Bardeen's method could be seen as using a special type of electromagnetism to explain where energy and momentum come from.  Since then, scientists have thoroughly studied these common solutions  \cite{Balart:2014cga, Toshmatov:2017zpr}, for black holes in many research papers. They've looked into things like quantum changes  \cite{Maluf:2018ksj}, how heat works, the effects of constants in space, and how quintessence plays a role \cite{Rodrigues:2022qdp,Maluf:2022qfc}.  To obtain a thorough comprehension of the matter source with nonlinear electrodynamics, there exists a review article that can be consulted for a comprehensive understanding of the subject matter \cite{Sorokin:2021tge}.

Modified gravity models offer an intriguing dynamic option to the $\Lambda$CDM cosmology by effectively explaining the current acceleration in the expansion of our Universe, known as the dark energy phase. Additionally, extensive research has been conducted on modified $f(R)$ gravity (refer to \cite{Sotiriou:2008rp,Nojiri:2010wj,DeFelice:2010aj} for a comprehensive review), which has consistently demonstrated its ability to successfully pass tests conducted within the solar system. {  Several sources have been investigated, notably \cite{Nojiri:2006ri,Capozziello:2003tk,Nojiri:2003ft,Nashed:2018oaf,Faraoni:2005vk}, covering topics such as cosmic acceleration and the investigating of the cosmological features of $f(R)$}. Other successful theories of gravity include additional modifications such as ${\textit f(R, T)}$ theory \cite{Harko:2011kv} and a newly suggested ${\textit f(R)}$ with scalar field  \cite{Kleidis:2018cdx}. These modified theories have the potential to effectively tackle the challenges posed by late-time cosmic acceleration by employing certain cosmological models. The present study aims to continue our studies in $f(R)$ gravitational theory to derive RBHs in the frame of $f(R)$ coupled with a scalar field.

The essence of the present study can be summarized as: In Section \ref{S2}, we present the theory of $f(R)$ with a scalar field. In Section \ref{S3}, we utilize the field equations to analyze a spherically symmetric spacetime, considering two distinct metric potentials, and deduce the resulting non-linear differential equations We solved these differential equations assuming equal metric potentials and unequal metric potentials. In Section \ref{S4}, we delve into the pertinent physics of the acquired solution in the scenario with differing metric potentials, as it presents the most intriguing case. In Section \ref{SeC.V}, our dedication lies in conducting a thorough examination of the black hole solutions that have been derived, with a particular focus on diverse elements such as the investigation of geodesics linked to these black holes. Our specific aim is to explore and analyze the paths taken by both particles moving through time and those moving at the speed of light within these specific spacetimes.  The concluding section is dedicated to providing a summary of the current investigation.
\section{$f(R)$ gravity coupled with scalar field}\label{S2}
In this section, we intend to introduce the fundamental framework of $f(R)$ with scalar-tensor gravity, and then employ their equations of motion to acquire a self-contained dynamical system.
The general form of an $f(R)$-gravity action, which includes a scalar  field  can be expressed as follows \cite{Kleidis:2018cdx}:
\begin{equation}\label{action}
S_{m}=\int d^{4}x \sqrt{-g} \left\{\frac{1}{\kappa^2}\left[f\left(R\right)+\psi(\chi)\chi_{;\alpha}\chi^{;\alpha}
\right]+\mathcal{L}_{m}\right\}dx\,,
\end{equation}
where $f$ is the unknown function of the Ricci scalar, $R$,  $\kappa^2=\frac{8\pi G}{c^4}$ where ${\mathrm G}$ refers to the  gravitational constant, ${ c}$ is the speed of light, and $\chi$ is the scalar field.
The ${L}_{m}$ is referred to as the matter Lagrangian density\footnote{In this study we are interested in vacuum solution so from now on we will omit the matter Lagrangian density.}, $\psi$ is a  scalar field $\chi$ and  $g$ represents the determinant of  $g_{\mu\nu}$.

$f(R)$ gravity possesses an additional scalar degree of freedom, expanding the range of possible solutions beyond what's typical in  GR  and $f(R)$ theories. This uniqueness places it in a distinct category from commonly studied modified gravity theories. By varying the action $S_m$ in Eq. (\ref{action}) w.r.t. the metric tensor, we derive the modified field equations.
\begin{eqnarray}\label{fe}
I_{\mu \nu}\equiv f_{R}R_{\mu\nu}-\frac{1}{2}\left[f+\psi(\chi) \chi_{;\alpha}
 \chi^{;\alpha}\right]g_{\mu\nu}-f_{R;\mu\nu}+g_{\mu\nu}\Box f_{R}+\psi(\chi)\chi_{;\mu}\chi_{;\nu}=0
\end{eqnarray}
where $\Box\equiv\nabla^{\mu}\nabla_{\mu}$.
Covariant derivatives contribute to the field equations becoming fourth-order partial differential equations. They simplify to ordinary $f(\mathcal{R})$ gravity equations when $\chi=0$ and to  GR equations when $f({R})={R}$ and $\chi=0$. Given that $f({R})$ is a multivariate analytic function, we adopt the notations $f({R})\equiv f$ and $f_{{R}}\equiv\frac{\partial f}{\partial{{R}}}$ for simplicity, as these derivatives are used extensively throughout this paper.

The variation of action (\ref{action}) w.r.t. the scalar field $\chi$ results in
\begin{equation}\label{4}
I_{Sc}\equiv \frac{1}{2}\psi'\chi_{;\mu}\chi^{;\mu}+\nabla_{\mu}(\psi\nabla^{\mu}\chi)=0\,.
\end{equation}
In the equation, the prime symbol denotes differentiation concerning the scalar field, expressed as $\psi'=\frac{d\psi}{d\chi}$. Equation (\ref{4}) essentially represents the Klein-Gordon equation.

Moreover, contracting the field equations (\ref{fe}), gives:
\begin{equation}\label{trace}
I\equiv \mathcal{R}f_{R}-2f+3\Box f_\mathcal{R}-\psi(\chi) \chi_{;\mu}\chi^{;\mu}.
\end{equation}
Using q. (\ref{trace}) we derive the form of  $ f(R)$ as:
\begin{eqnarray} \label{f3s}
f(R)=\frac{1}{2}\big[3\Box {f_{R}}+{ R}{f_{R}}-\psi(\chi) \chi_{;\mu}\chi^{;\mu}\Big]\,.\end{eqnarray}
Using Eq. (\ref{f3s}) in Eq. (\ref{fe}) we get \cite{Kalita:2019xjq}
\begin{eqnarray} \label{f3ss}
{\mathop{\mathcal{ I}}}_{\mu \nu}={ R}_{\mu \nu} {f_{R}}-\frac{1}{4}g_{\mu \nu}{ R}{ f}_{{ R}}+\frac{1}{4}g_{\mu \nu}\Box{ f}_{{ R}} -\nabla_\mu \nabla_\nu\mathit{ f}_{{ R}}-\frac{1}{4}\psi(\chi) \chi_{;\alpha}\chi^{;\alpha}g_{\mu \nu}+\psi(\chi) \chi_{;\mu}\chi^{;\nu}=0  \,.\end{eqnarray}
Therefore, it is crucial to analyze Equations (\ref{4}), (\ref{trace}), and (\ref{f3ss}) in the context of a spherically symmetric spacetime characterized by two unknown functions.


\section{Regular Black Hole Solutions}\label{S3}
This section introduces a four-dimensional solution for a black hole in $f(R)$ theory connected to a scalar field. We assume the spacetime configuration is described by the metric
\begin{equation} \label{m2}
ds^2= -h(r)dt^2+\frac{1}{h(r)\,h_1(r)}dr^2+r^2d\Omega^2\,, \qquad \mbox {where} \qquad d\Omega^2=d\theta^2+\sin^2\theta d\phi^2\,,\end{equation}
is the two-dimensional sphere\footnote{In this study we are interested in the study of spherically symmetric cases where the metric potentials depend on the radial coordinate. Consequently, we will assume the form of $f(R)=f(R(r))\equiv f(r)$.}.   Here $h(r)$ and $h_1$ are unknown functions that depend solely on the radial coordinate $r$ because our interest lies in the spherically symmetric case. For the spacetime (\ref{m2}), we get the Ricci scalar
\begin{equation}\label{ri}
R=-\frac{2r^2 h''h_1+r^2h'h'_1+8rh'h_1+4rhh'_1-4(1-h\,h_1)}{2r^2}.
\end{equation}
Then, the non-vanishing components of Eqs. (\ref{4}), (\ref{trace}) and (\ref{f3ss}), when $T_{\mu \nu}=0$, read the following set of field equations:
\begin{eqnarray} \label{fe13}
& & I_t{}^t=\frac {2\,F''    h  h_1  {r}^{2}-2\,F   h'' h_1  {r}^{2}+r \left[ rF' h  -F  \left(  h' r-4\,h   \right)  \right] h'_1 -2\,rh_1   \left[  h'r-2\,h   \right] F'  +4\,F  h  h_1  -4\,F  -2\,\psi  h  h_1   \chi'^{2}{r}^{2}}{8{r}^{2}}\equiv 0,\nonumber\\
& & I_r{}^r= \frac {4\,F  h  h_1-2\,F   h''  h_1  {r}^{2}-6\,F''h h_1 {r}^{2}- \left[ 3\, F'h  {r}^{2}+rF   \left(  h' r+ 4\,h   \right)  \right] h'_1 -2\,rh_1   \left(  h' r-2\,h   \right) F'    -4\,F  +6\,\psi h  h_1  \chi'^{2}{r}^{2}}{8{r}^{2}}\equiv 0,\nonumber\\
& & I_{\theta}{}^{\theta} =I_{\phi}{}^{\phi} =\frac {2\,F   h'' h_1  {r}^{2}+2\, F'' h h_1  {r}^{2}+r \left( 2\,h_1  rh'  +h   \left(rh'_1 -4\,h_1    \right) \right) F'  +4\,F  -4\, F  h  h_1 +F   h'  h'_1 {r}^{2}-2\,\psi  h  h_1  \chi'^{2}{r}^{2}}{8{r}^{2}}\equiv 0\,,\nonumber\\
& & I_{Sc}=\frac { \left\{  \psi' h  h_1  \chi' r-2\, \psi' h  h_1  r-\psi   \left[ 2\,h_1  rh'  +h   \left( rh'_1  +4\,h_1   \right)  \right] \right\} {\chi' }-2\,\psi  h  h_1   \chi'' r}{2r}\equiv 0\,,\nonumber\\
& & I=\frac {1}{2{r}^{2}}\bigg\{6 F'' h  h_1  {r}^{2}-2F   h'' h_1  {r}^{2}+ \left[2(3F' r-4F)  rh_1  -F  h'_1 {r}^{2}  \right]h'  +\left(3F' r-4F   \right)  rh h'_1  +4F  -4 F  h  h_1  -2\psi  h  h_1   \chi'^{2}{r}^{2}\nonumber\\
& &+12 F'h  h_1  r-4f  {r}^{2}\bigg\}\equiv0\,,
\end{eqnarray}
{ where $F=f_R=\frac{df(R)}{dR}=$$\frac{df(R(r))}{dr} \times \frac{dr}{dR}$, $h'=\frac{dh(r)}{dr}$, $h'_1=\frac{dh_1(r)}{dr}$, and $F'=\frac{dF(r)}{dr}$}\footnote{Since we are studying a spherically symmetric case in which the unknown functions depend $r$ then we will assume that $f(R)\equiv f(R(r))$ and then its derivatives $F=f_R=\frac{df(R(r))}{dR}=\frac{df(R(r))}{dr} \times \frac{dr}{dR}$.}.  We will solve the field equations (\ref{fe13}) using the constraints  $h_1(r)=1$ and $h_1(r)\neq 1$:\\
\subsection{The case with equal metric, i.e., $h_1$(r)=1 ``Hayward solution'' }
{ In this scenario, if we suppose $h(r)$ to take the form:
\begin{align}\label{sols}
h(r)=1-\frac{2Mr^2}{r^3+g^3}\;,
\end{align}
 where $g$ is a dimensional constant that  has the unit of ${\textrm length}$.
If we use Eq. (\ref{sols}) in the first three differential of Eq.~(\ref{fe13}) we get the form of $F(r)$ and the function $\psi(\chi)$ in the form:}
\begin{align} \label{so1}
&  F(r)=\frac{c_1\chi'^{2} \left( r+g \right)^ {2} \left( {r}^{2}-rg+{g}^{2} \right)^{2} \left( 3{r}^{5}M-{g}^{6}-2{g}^{3}{ r}^{3}-{r}^{6} \right) ^{2}}{18{g}^{3}{M}{r}^{3} \left( 4{g}^{9}-5{r}^{9}+12M{r}^{8}-6{r}^{6}{g}^{3}+21M{r}^{5}{g}^{3} +3{r}^{3}{g}^{6} \right) }
\nonumber\\
&\times \exp\bigg\{2\int \bigg[15 {r}^{15}\chi'  -15\chi' {g}^{12}{r}^{3}+5 \chi''{r}^{16}-66\chi' {g}^{9}{r}^{6}-48\chi'{g}^{6}{r}^{9}+90{r}^{13} \chi'{M}^{2}-72{r}^{14} \chi' M-45{g}^{6} \chi'' {r}^{9}M-63{g}^{3} \chi'' {r}^{12}M\nonumber\\
&+63{g}^{3} \chi''{M}^{2}{r}^{11}+36 \chi'' {M}^{2}{r}^{14}-27\chi'' {r}^{15}M-9{g}^{9}{r}^{6} \chi''M+14{g}^{6} \chi'' {r}^{10}-4{g}^{9}\chi''{r}^{7}+90 \chi'{g}^{6}{r}^{8}M-144\chi'{g}^{3}{r}^{11}M+6 \chi' {g}^{15}\nonumber\\
&+252\chi' {g}^{3}{r} ^{10}{M}^{2}+12 \chi' {g}^{3}{r}^{12}-11{g}^{12} \chi'' {r}^{4}+162 \chi' {g}^{9}{r}^{5}M+16{g}^{3} \chi''{r}^{13}-4{g}^{15}r\chi'' \bigg]\bigg\{ \chi' r \left( {g}^{6}+2{g}^{3}{r}^{3}+{r}^{6}-3{r}^{5}M \right)\nonumber\\
&  \left( 12M{r}^{8}-5{r}^{9}-6{r}^{6}{g}^{3}+21M{r}^{ 5}{g}^{3}+3{r}^{3}{g}^{6}+4{g}^{9} \right) \bigg\}^{-1}{dr}\bigg\} \,,\nonumber\\
& \psi(r)=c_1exp\bigg\{2\,\int \bigg[15\,{r}^{15}\chi' -15\, \chi'  {g}^{12}{r}^{3}+5\, \chi' {r}^{16}- 66\,\chi'  {g}^{9 }{r}^{6}-48\,\chi' {g}^{6}{r}^{9}+90\,{r}^{13} \chi' {M}^{2}-72\,{r}^{14} \chi' M-45\,{g}^{6} \chi'' {r}^{9}M\nonumber\\
&-63\,{g}^{3} \chi'' {r}^{12}M+63\,{g}^{3} \chi'' {M}^{2}{r}^{11}+36\, \chi'' {M}^{2}{r}^{14}-27\, \chi''{r}^{15}M-9\,{g}^{9}{r}^{6} \chi'' M+14\,{g}^{6} \chi'' {r}^{10}-4\,{g}^{9} \chi'' {r}^{7}+ 90\, \chi' {g}^{6 }{r}^{8}M\nonumber\\
&-144\, \chi' {g}^{3}{r}^{11}M+6\, \chi' {g}^{15}+252\, \chi'{g}^{3}{r}^{10}{M}^{2}+12\, \chi' {g}^{3}{r}^{12}-11\,{g}^{ 12} \chi'' {r}^{4}+162\, \chi' {g}^{9}{r}^{5}M+16\,{g}^{3} \chi''{r}^{13}-4\,{g}^{15}r{ \frac {d^{2}}{d{r}^{2}}}\chi \bigg]\nonumber\\
&\bigg\{ \chi' r \left( {g}^{6}+2\,{g}^{ 3}{r}^{3}+{r}^{6}-3\,{r}^{5}M \right)  \left(12\,M{r}^{8} -5\,{r}^{9} -6\,{r}^{6}{g}^{3}+21\,M{r}^{5}{g}^{3}+3\,{r}^{3}{g}^{6}+4\,{g}^{9} \right) \bigg\}^{-1}{dr}\bigg\}\,.
 \end{align}
{ Using Eq.~(\ref{so1}) in the fourth equation of Eq. (\ref{fe13}) we get the scalar field $\chi$ given by Eq. ( 1) which is presented in the notebook of this study.   The exact solution of the differential equation  ( 1) which is presented in the notebook is difficult to derive however, we can find its asymptote solution as $r \to 0$ in the form:}
  \begin{align}\label{solsc}
  \chi(r)=c_2+\frac{10}{3} r\,.
  \end{align}
 We will now examine the scenario in which $h_1\neq 1$.
\subsection{ $h_1(r)\neq 1$ ``general solution: generalization of Hayward black hole solution''}
{  If we assume the components of the metric tensor $h(r)$ and $h_1(r)$  to have the form\footnote{The assumption of the metric potentials as given by (\ref{sol22}) ensures that the metric is regular, and we aim to derive their corresponding forms of $F(r)$, $\psi(r)$, and $\chi$, as we will show.}:
\begin{align}\label{sol22}
h(r)=1-\frac{2Mr^2}{r^3+g^3}\,, \qquad h_1(r)=e^{\alpha r^3}\,,
\end{align}
where $\alpha$ is a dimensional parameter has the unit of ${\textrm length^{-3}}$. If we substitute Eq. (\ref{sol22}) in Eq. (\ref{fe13}) we get the other unknowns, $F(r)$, $\chi$ and $\psi$ which are very lengthy so we list them in Section II of the notebook associated with this study. From these equations listed in Section II presented in the notebook we can show that when $\alpha=0$,   we return to the case $h_1=1$. The notebook's solution for the case with unequal metric has the following asymptotic form for $F(r)$:

\begin{align}\label{fa}
F(r)\approx -\frac{18c_2}{\alpha}+15c_2r^3-\frac{324 c_2M}{5\alpha g^6} r^5-\frac{341}{4}c_2\alpha r^6+\frac{c_2(43740M+2870M\alpha g^3)}{360\alpha g^9}\,.
\end{align}
In the upcoming section, we will discuss the physics associated with the case where $h_1 \neq 1$ because it represents the general scenario.}
 \section{Physical characterization of the case $h_1 \ne 1$}\label{S4}
We are going to explore the fundamental physics behind the solution (\ref{sol22}) and its related quantities, namely $F(r)$, $\psi(r)$, and $\chi(r)$. The physics of the black hole solution described by Eq. (\ref{so1}) has been extensively studied in the literature \cite{Bonanno:2022jjp,Cadoni:2022chn,Yang:2022fwp}.
\subsection{The relevant physics and the thermodynamics of the BH (\ref{sol22})}
\begin{figure*}
\centering
\subfigure[~The plot of the temporal component of the metric given by Eq.~(\ref{sol22}) viz. the radial coordinate $r$ when  $g=0.7$. ]{\label{fig:1a}\includegraphics[scale=0.25]{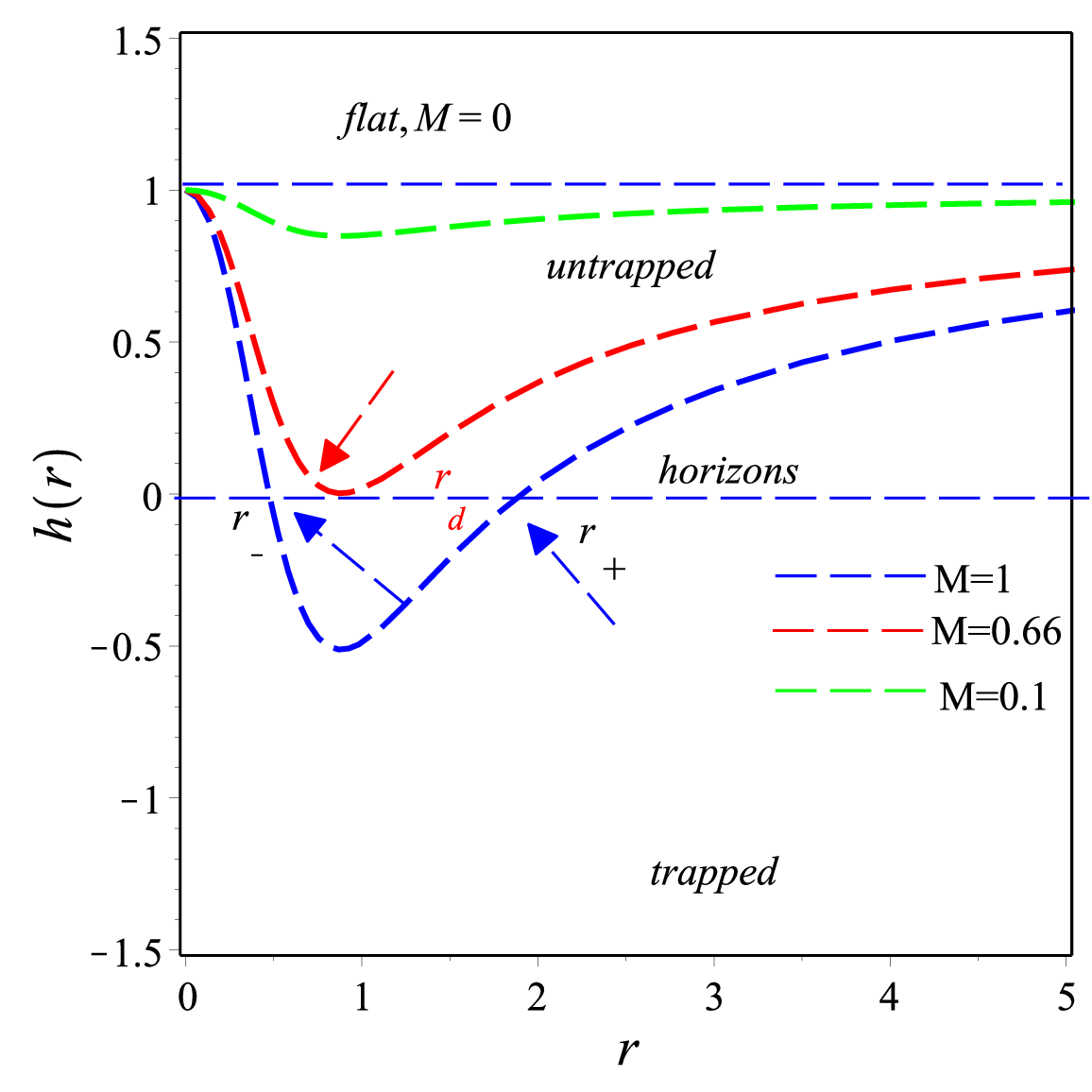}}\hspace{0.2cm}
\subfigure[~The graph of $g_{rr}$ given by Eq.~(\ref{sol22}) viz.   $r$. ]{\label{fig:1b}\includegraphics[scale=0.25]{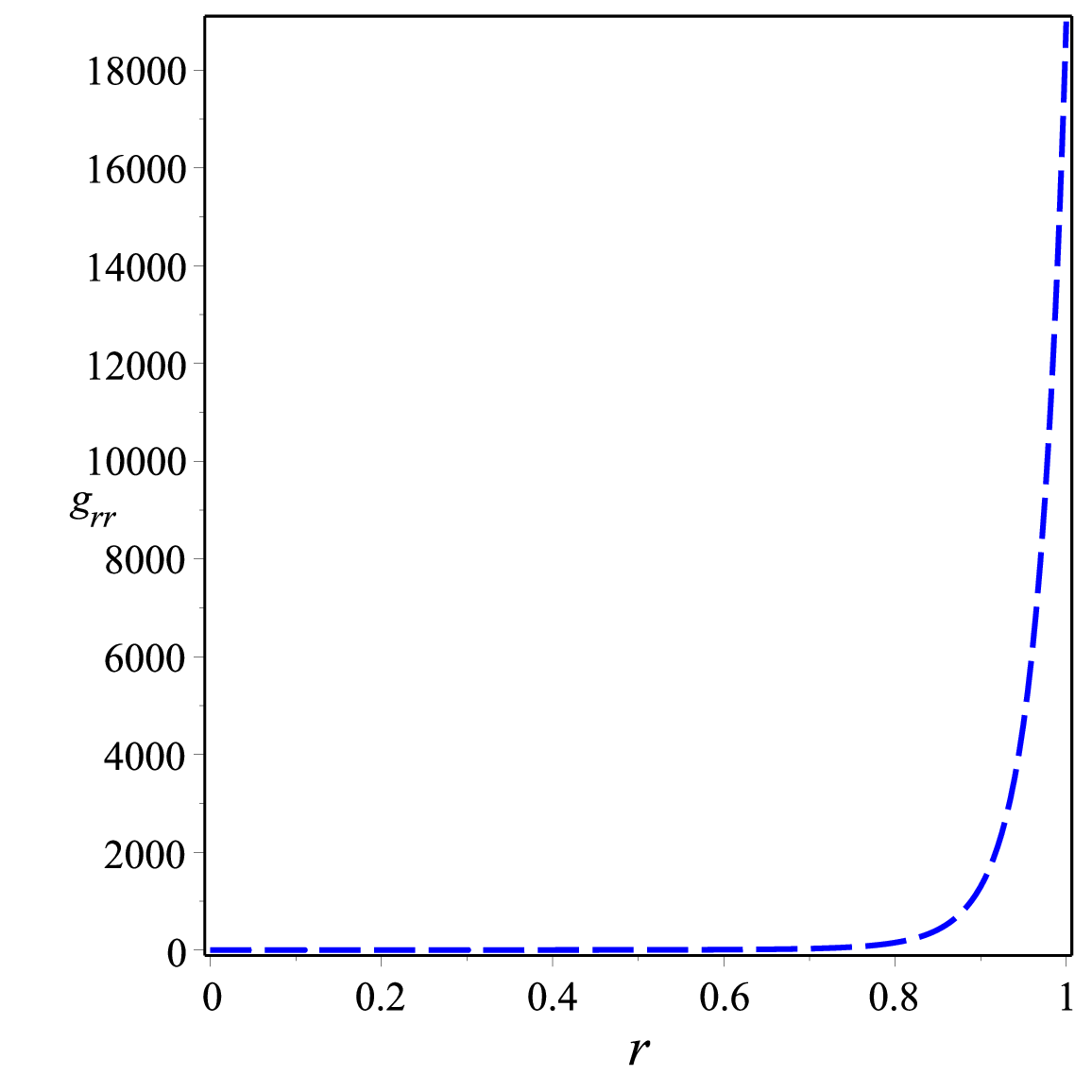}}\hspace{0.2cm}
\subfigure[~The graph of the  function $\psi$ and the scalar field $\chi$ of solution (\ref{so1}) viz. the radial coordinate $r$. ]{\label{fig:1c}\includegraphics[scale=0.25]{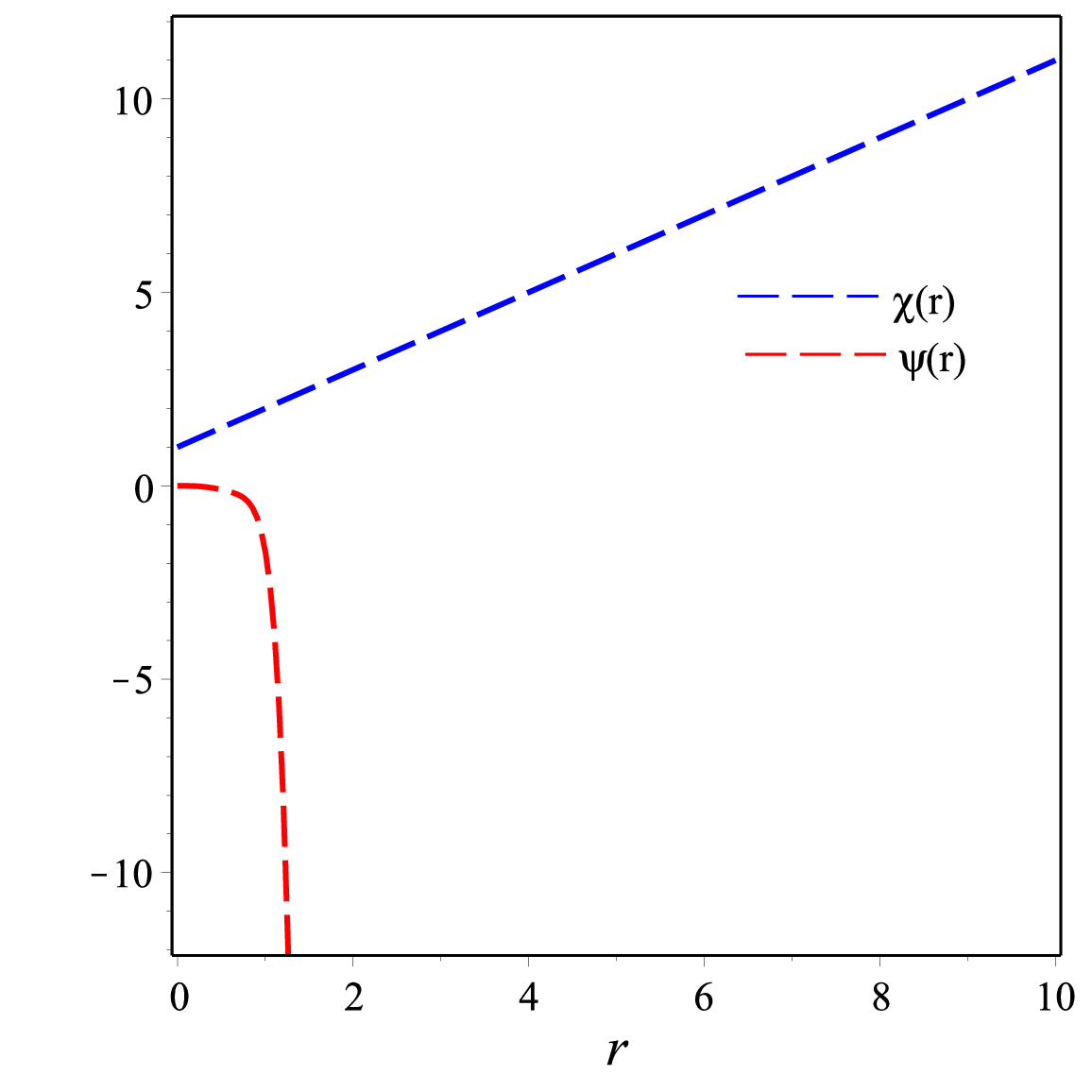}}\hspace{0.2cm}
\subfigure[~The plot of the  function $\psi$ and the scalar field $\chi$ of solution (\ref{sol22}) viz. the radial coordinate $r$. ]{\label{fig:1c1}\includegraphics[scale=0.25]{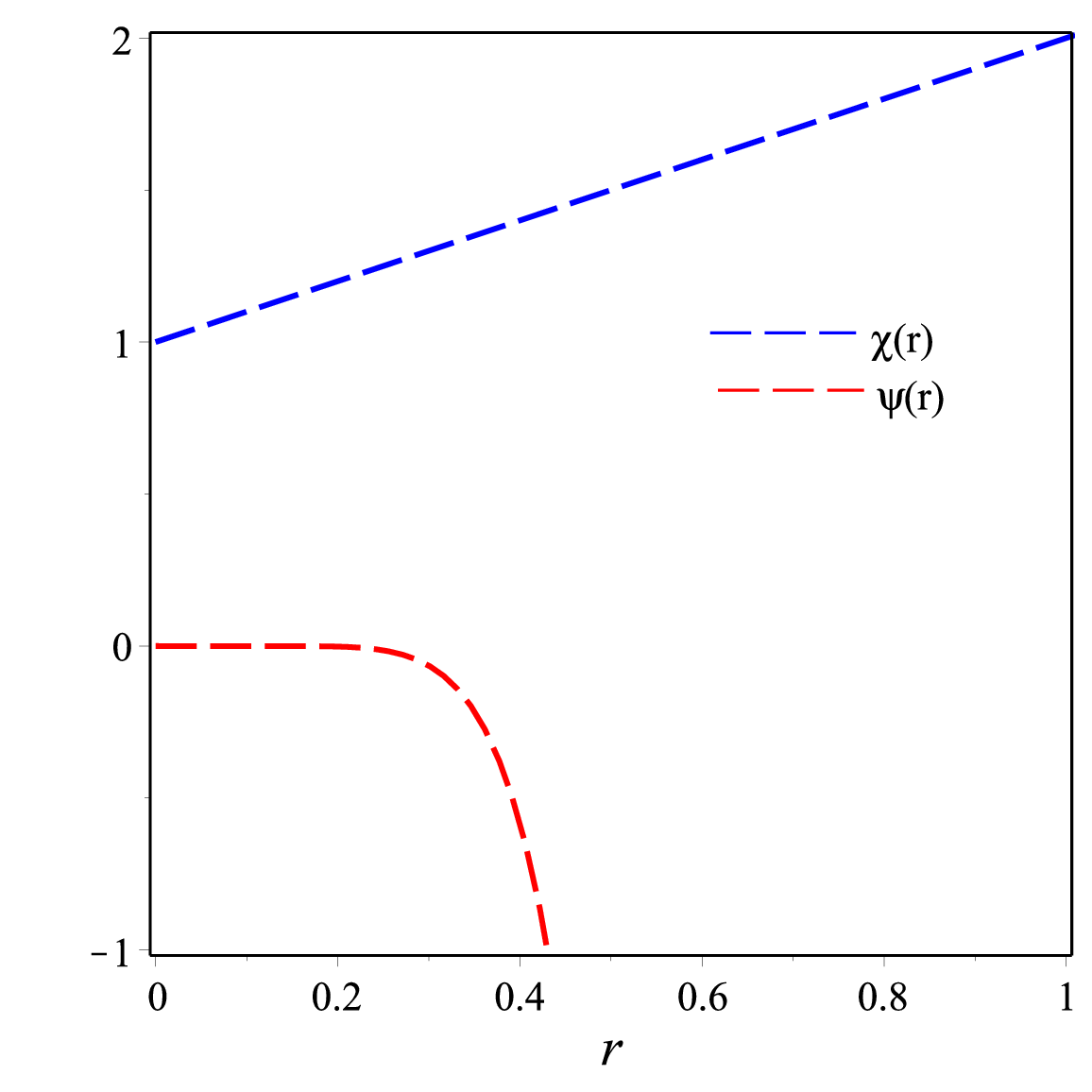}}\hspace{0.2cm}
\subfigure[~The Ricci scalar of the solution (\ref{so1}) viz the radial coordinate $r$.  ]{\label{fig:1d}\includegraphics[scale=0.25]{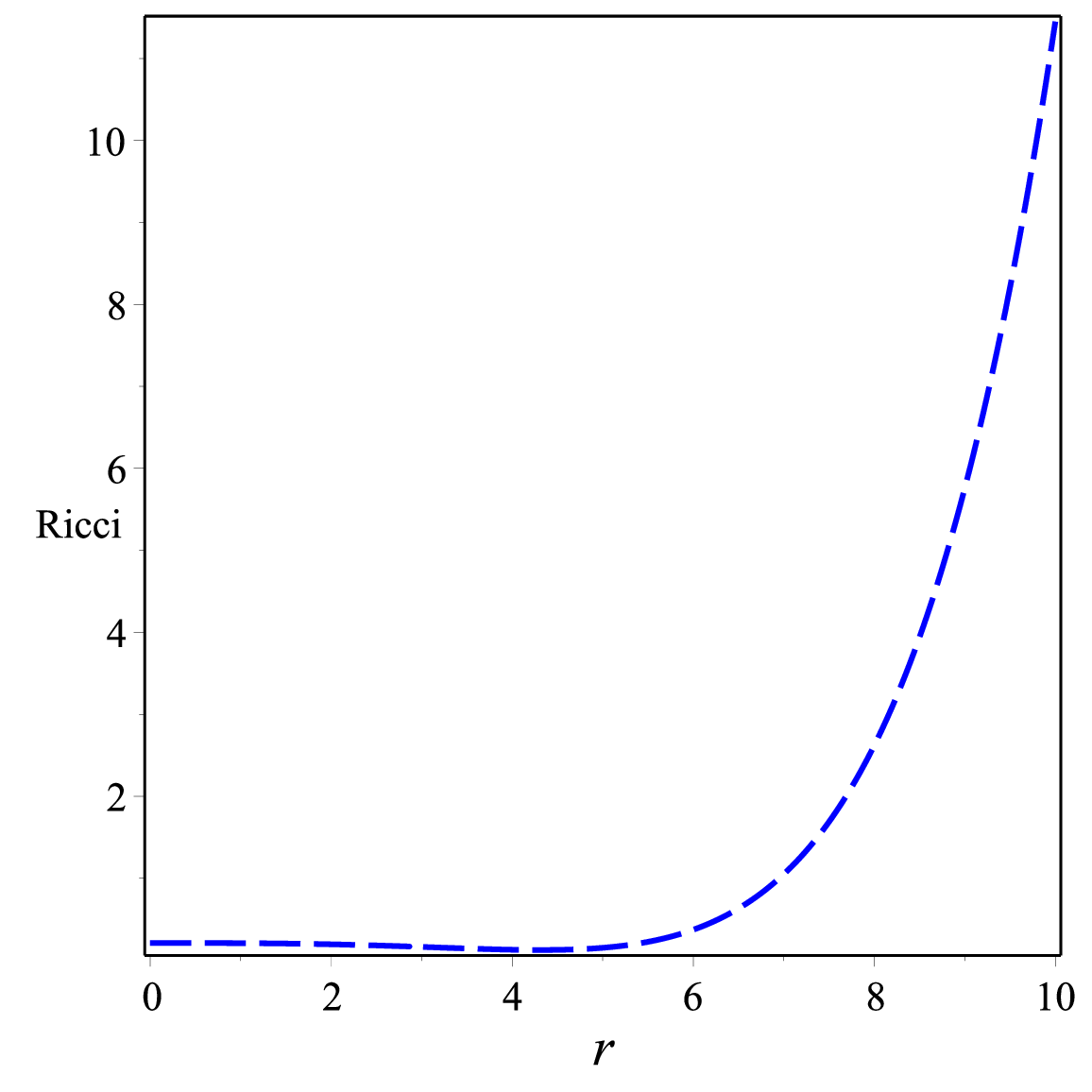}}
\subfigure[~The Ricci scalar of the solution (\ref{sol22}) viz the radial coordinate $r$. ]{\label{fig:1e}\includegraphics[scale=0.25]{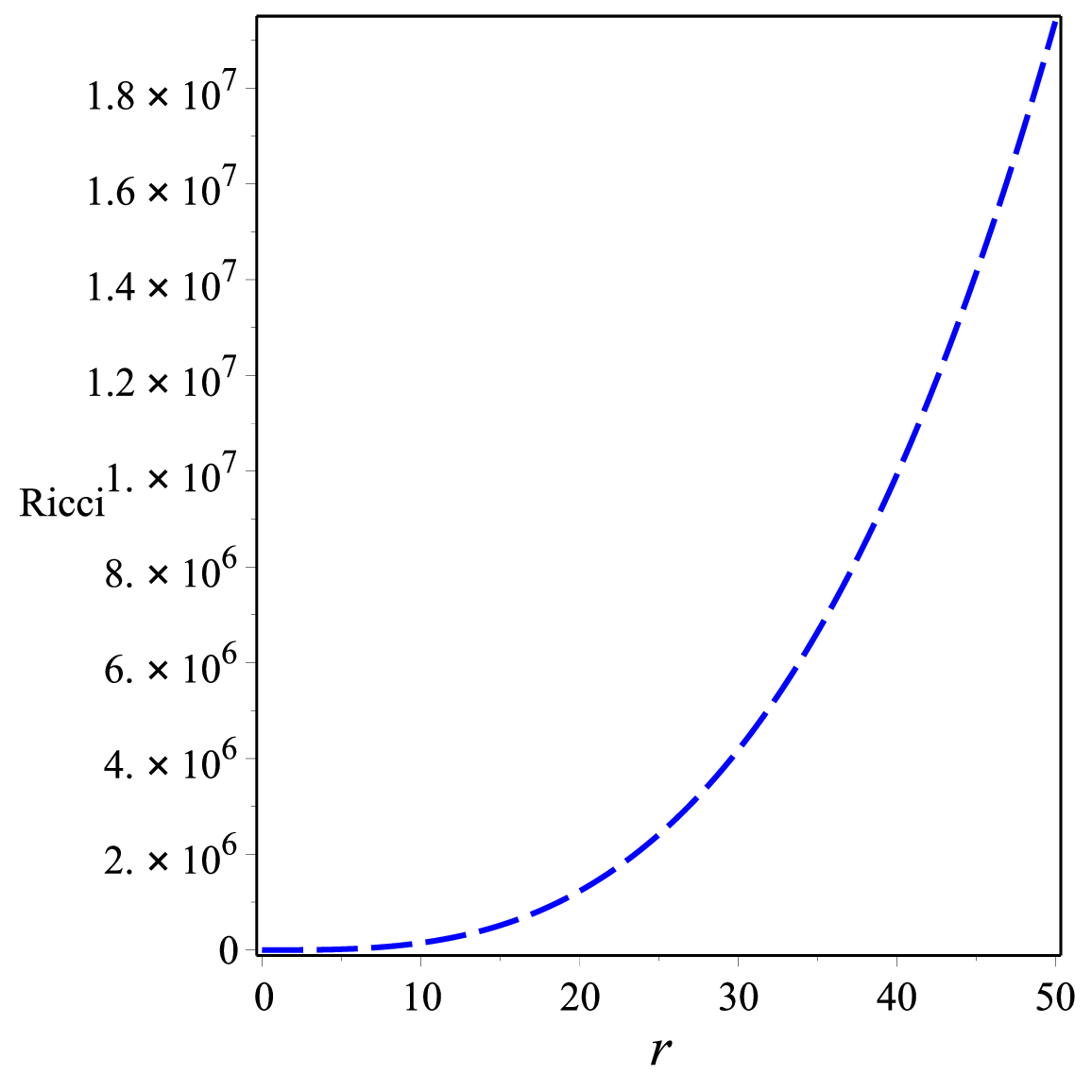}}
\caption{Schematic representations of  the coordinate $r$ \subref{fig:1a} viz. the  temporal component of the metric  given by Eq.~(\ref{mete}) which indicates the horizons, $r_-$ and $r_+$, the degenerate horizon $r_d$ and the naked singularity region;  \subref{fig:1b} vs. the $g_{tt}$ given by Eq.~(\ref{sol22}); \subref{fig:1c} the function $\psi$ and the scalar field $\chi$; \subref{fig:1d} and \subref{fig:1e}  represent the Ricci
scalar of the solutions (\ref{so1}) and (\ref{sol22}). All the plots are drawn when $M=10$, $g=100$, $c_2=-1$ for Eq. (\ref{so1}) while for Eq. (\ref{sol22})  we put $M=10$, $g=100$, $c_2=-0.01$, and  $\alpha=100$}
\label{Fig:1}
\end{figure*}

To extract the physics of the case $h_1 \neq 1$ we first are going to write its line element which takes the form:
\begin{align}\label{mete}
ds^2=-\left(1-\frac{2Mr^2}{r^3+g^3}\right)dt^2+\frac{dr^2}{\left(1-\frac{2Mr^2}{r^3+g^3}\right)e^{\alpha r^3}}+d\Omega^2\,.\end{align}
The above metric has the following form as $r \to0$
\begin{align}\label{meta}
ds^2\approx -\left(1-\frac{2Mr^2}{g^3}+\frac{2Mr^5}{g^6}\right)dt^2+\frac{dr^2}{\left(1-\frac{2Mr^2}{g^3}+\alpha r^3+\frac{2Mr^5(g^3-\alpha)}{g^6}\right)}+d\Omega^2\,.\end{align}
The line element (\ref{mete}) reduces to the  Hayward solution when the dimensional parameter $\alpha=0$ and reduces to the Schwarzschild black hole when the dimensional parameters $\alpha=g=0$.  The metric's (\ref{mete}) behavior is illustrated in Fig.\ref{Fig:1} \subref{fig:1a}.

\begin{figure*}
\centering
\subfigure[~The functions $f(R)$ viz  $R$ ]{\label{fig:2a}\includegraphics[scale=0.25]{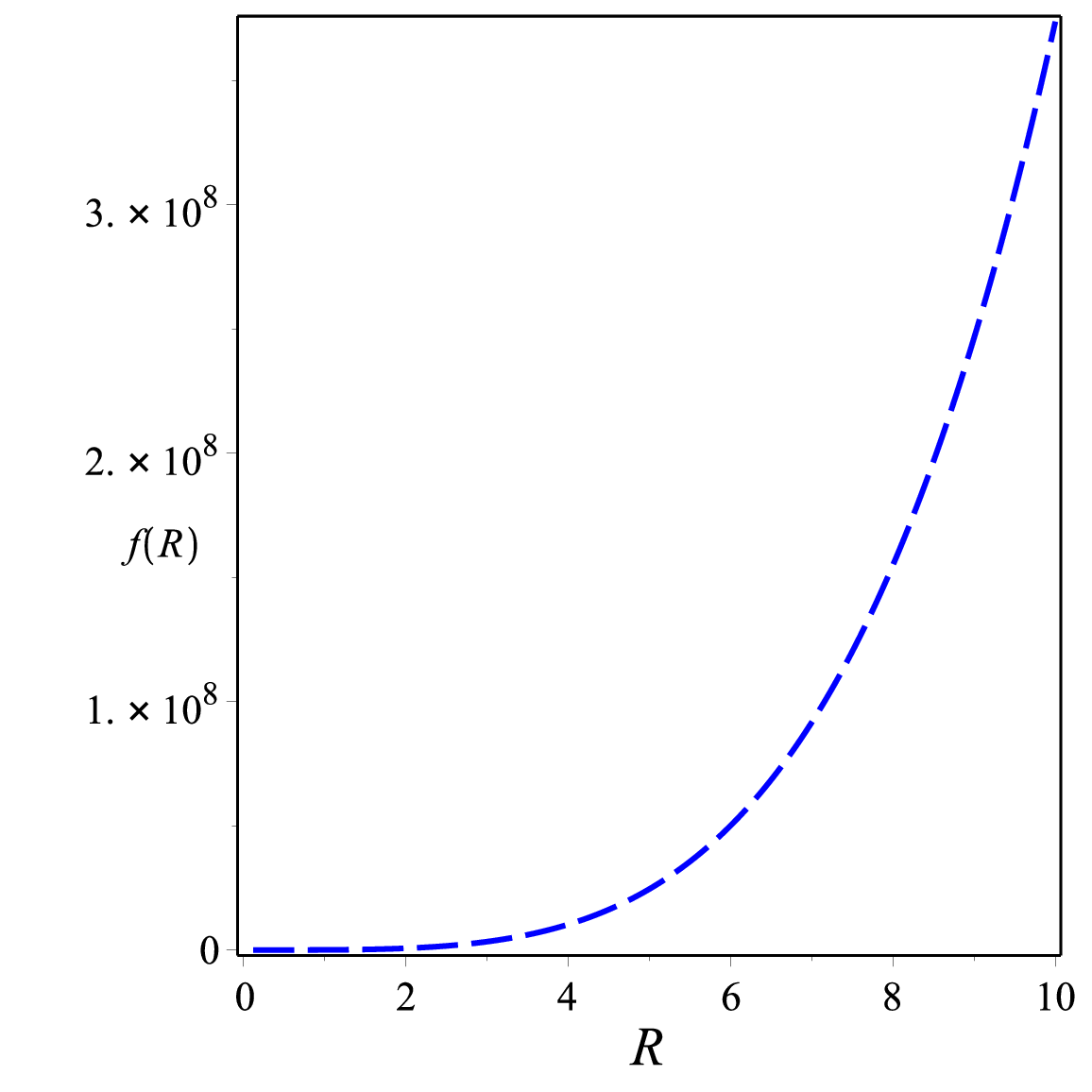}}
\subfigure[~$f'(R)$ viz  $R$ ]{\label{fig:2b}\includegraphics[scale=0.25]{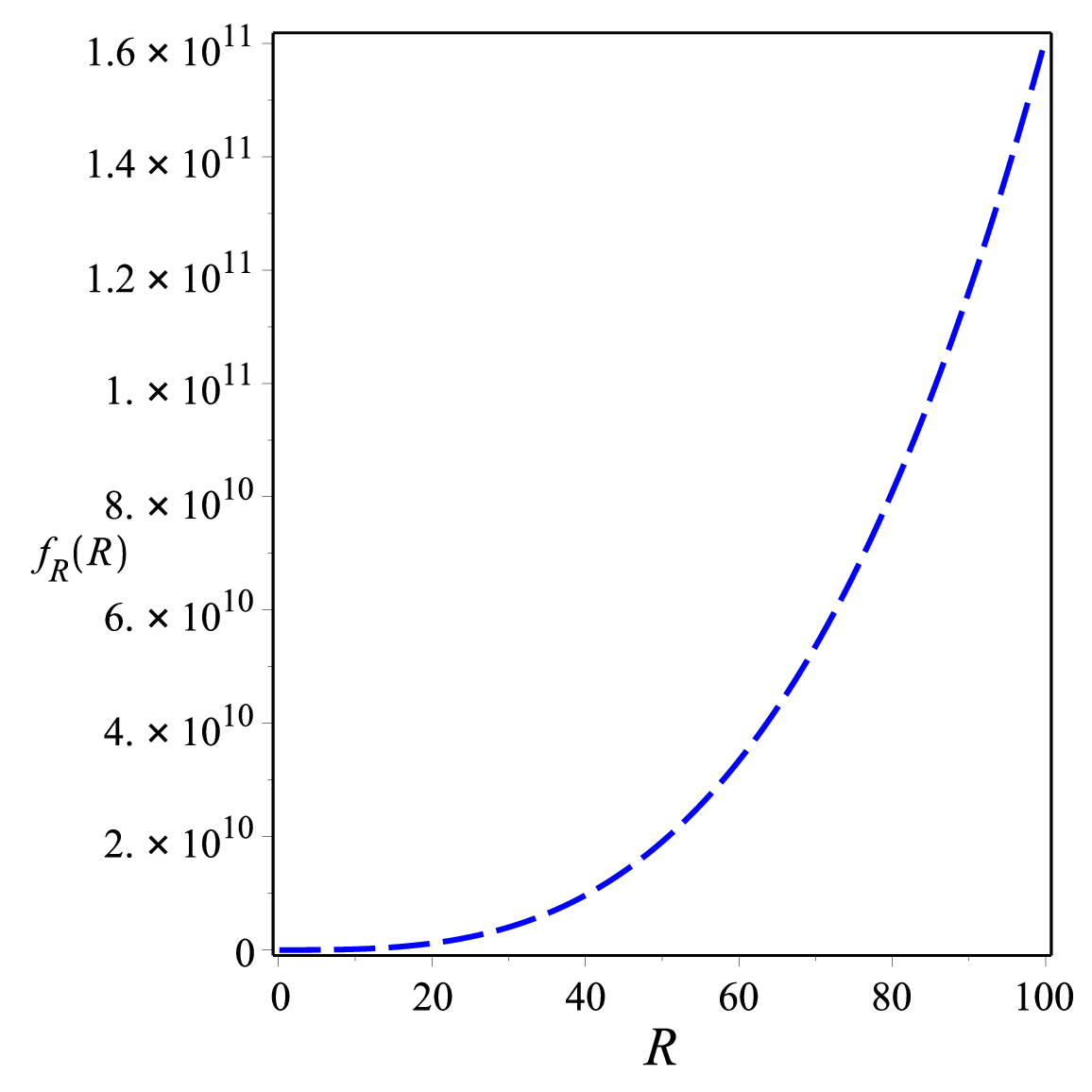}}
\subfigure[~$f''(R)$ viz  $R$  ]{\label{fig:2c}\includegraphics[scale=0.25]{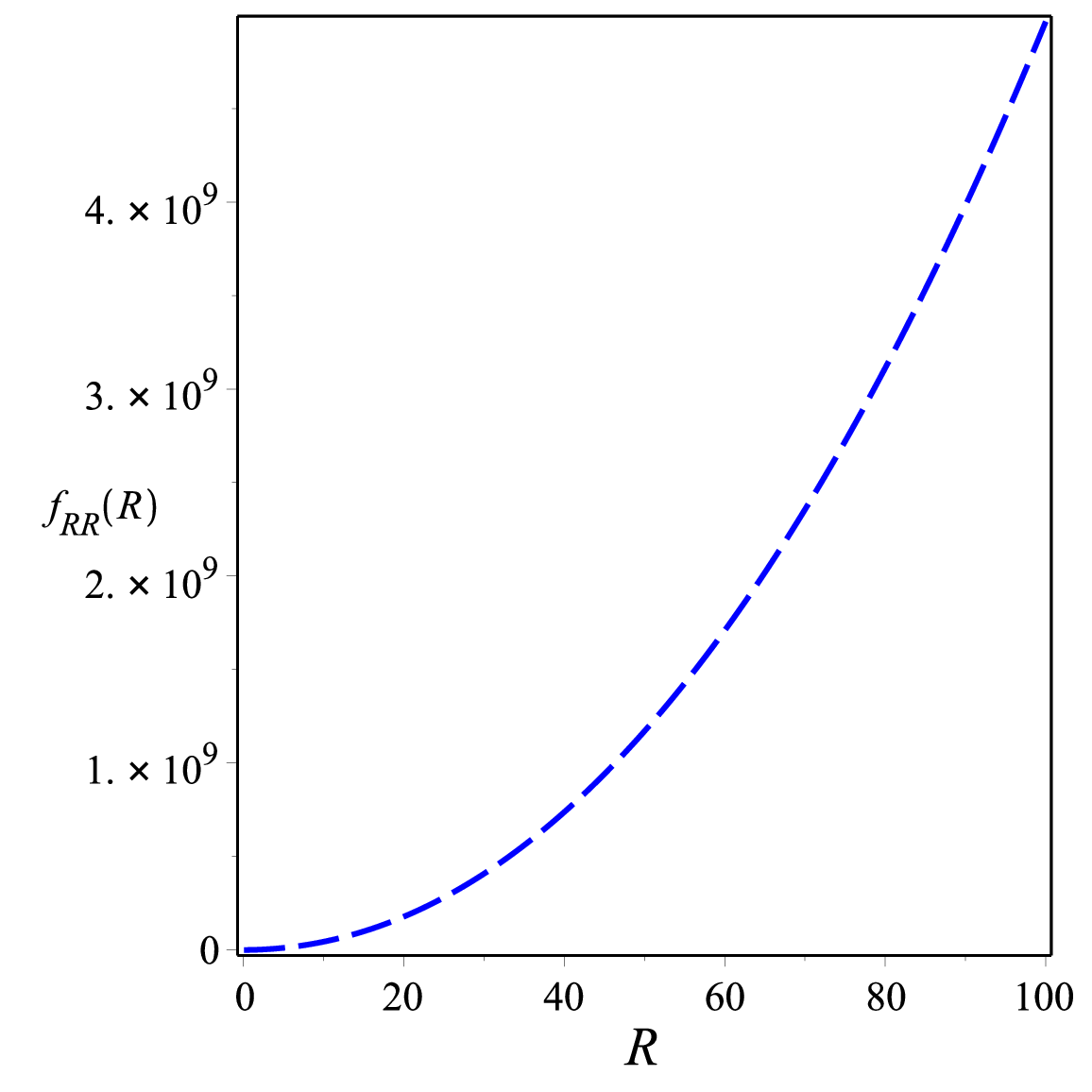}}
\caption{Schematic diagrams of  \subref{fig:2a} $f(R)$;  \subref{fig:2b} $f_R(R)$; \subref{fig:2c} $f_{RR}(R)$ viz Ricci scalar $R$ for the solution (\ref{so1}). We used the following numerical values to reproduce the above diagrams: M = 10, $M=10$ and $g=100$ and $c_1=-1$}
\label{Fig:2}
\end{figure*}

As shown in  Eqs. (\ref{mete}) or (\ref{meta}) that as $r\to 0$ we have finite values for $g_{tt}$ and $g_{rr}$ which ensure the regularity of this black hole.
As we stated the solution given by Eq. (\ref{so1}) is known in GR here we will list its Ricci scalar because we need it to write the form of $f(R)$ associated with it. The Ricci scalar of solution (\ref{so1}) takes the form:
\newpage
\begin{align}\label{ricci1}
R=\frac{24M\left(1-\frac{r^3}{2g^3}\right)}{g^3\left(1+\frac{r^3}{g^3}\right)^3}\approx \frac{24M}{g^3}-\frac{84M}{g^6}r^3 \quad {\mbox as} \quad  r\to 0.
\end{align}
{  Using the solution given by Eq. (\ref{so1})  we can get the form of $f(r)$ using the relation: footnote{The form of $f(r)$ can be calculated from Eq.~(\ref{f3s}) by substituting Eq.~(\ref{so1}). By employing the Ricci scalar's form as provided in Equation (\ref{ricci1}) in terms of the function $f(r)$, we can deduce the form of $f(R)$ as expressed in Equation (\ref{f1R}).} \begin{align}\label{fr} f(r)=\int f_R(r) dR(r)=\int F(r) \frac{dR}{dr} dr\,.\end{align} Using the value of$R$ presented in  Eq. (\ref{ricci1}) in $f(r)$ given by Eq. (\ref{fr}) we get:}
 \begin{align}\label{f1R}
&f(R)\approx{\frac {65{M}^{5}c_1}{43218{g}^{5}}}
 \left( {98}^{2/3}{M}^{5}+{\frac {12005}{117}}\,g{M}^{4}
 \right) R-{\frac {25}{1555848}}\,{\frac {c_1\,
{g}^{8}{98}^{2/3}{R}^{2}}{M}}-{\frac {25}{21003948}}\,{\frac {c_1\,{98}^{2/3}{g}^{11
}{R}^{3}}{{M}^{2}}}\nonumber\\
 &{\frac {925}{
48393096192}}\,{\frac {c_1\,{g}^{14}{98}^{2/3}{R}^{4}}{{
M}^{3}}}+{\frac {145}{1742151462912}}\,{\frac {c_1\,{g}^{17}{98}^{2/3}{R}^{5}}{{M}^{4}}}+{\frac {395}{376304715988992}}\,{\frac {c_1{g}^{20}{98}^{2/3}
{R}^{6}}{{M}^{5}}}\,.\nonumber\\
 \end{align}
 Eq. (\ref{f1R}) demonstrates unequivocally that the ongoing $c_1$ has a dimensional quantity of order ${\textrm (lenght)^{-5}}$.
Using Eq. (\ref{f1R}), it is easy to find the first and second derivatives of $f(R)$ w.r.t $R$  to show if the model under consideration is stable or not. We show this behavior in Fig. \ref{Fig:2}. The plots in Figure \ref{Fig:2} demonstrate that the Hayward solution in the frame of $f(R)$ is stable due to the conditions $f(R)>0$, $f_R(R)>0$, and $f_{RR}(R)>0$.
\newpage
{ Next, we are going to calculate the invariants of the line element (\ref{mete}) and we list them in the notebook given by Eqs. ({ 7}),   ({ 8}), and  ({ 9}). The asymptotic form of these invariants, listed in the notebook given by Eqs. ({ 7}),  ({ 8}), and  ({ 9})  as $r\to 0$ takes the form:}
\begin{align}
&K=R_{\mu \nu \rho \sigma}R^{\mu \nu \rho \sigma} \approx \frac{96M^2}{g^6}-\frac{64\alpha M}{g^3}r+22\alpha^2r^2\,,\nonumber\\
&R_{\mu \nu}R^{\mu \nu} \approx \frac{144M^2}{g^6}-\frac{96\alpha M}{g^3}r+43\alpha^2r^2\,,\nonumber\\
&R\approx \frac{24M}{g^3}-8\alpha r-\frac{42M(2-\alpha g^3)}{g^6}r^3\,.
\end{align}

\begin{figure*}
\centering
\subfigure[~The functions $f(R)$ viz  $R$ ]{\label{fig:3a}\includegraphics[scale=0.25]{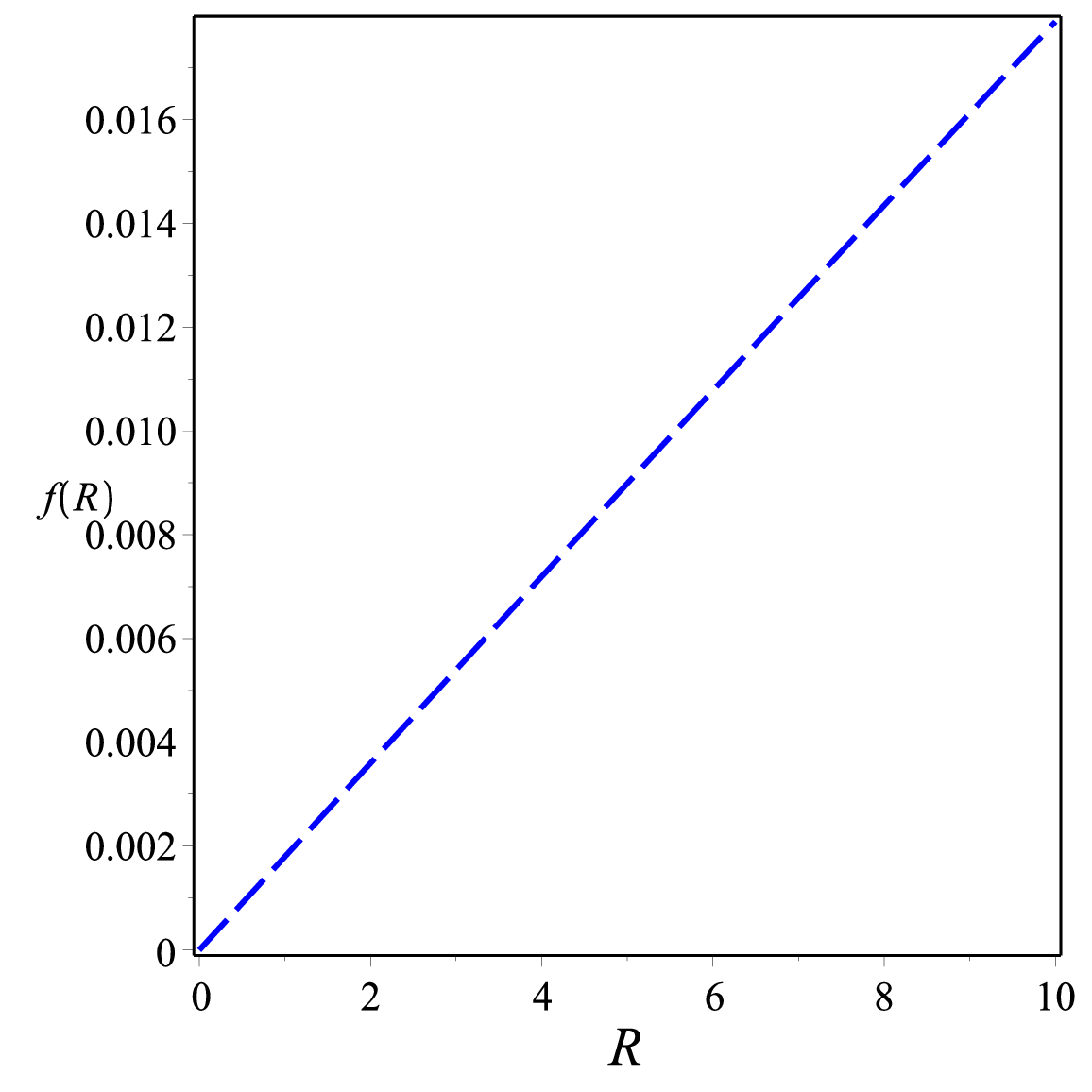}}
\subfigure[~$f'(R)$ viz  $R$ ]{\label{fig:3b}\includegraphics[scale=0.25]{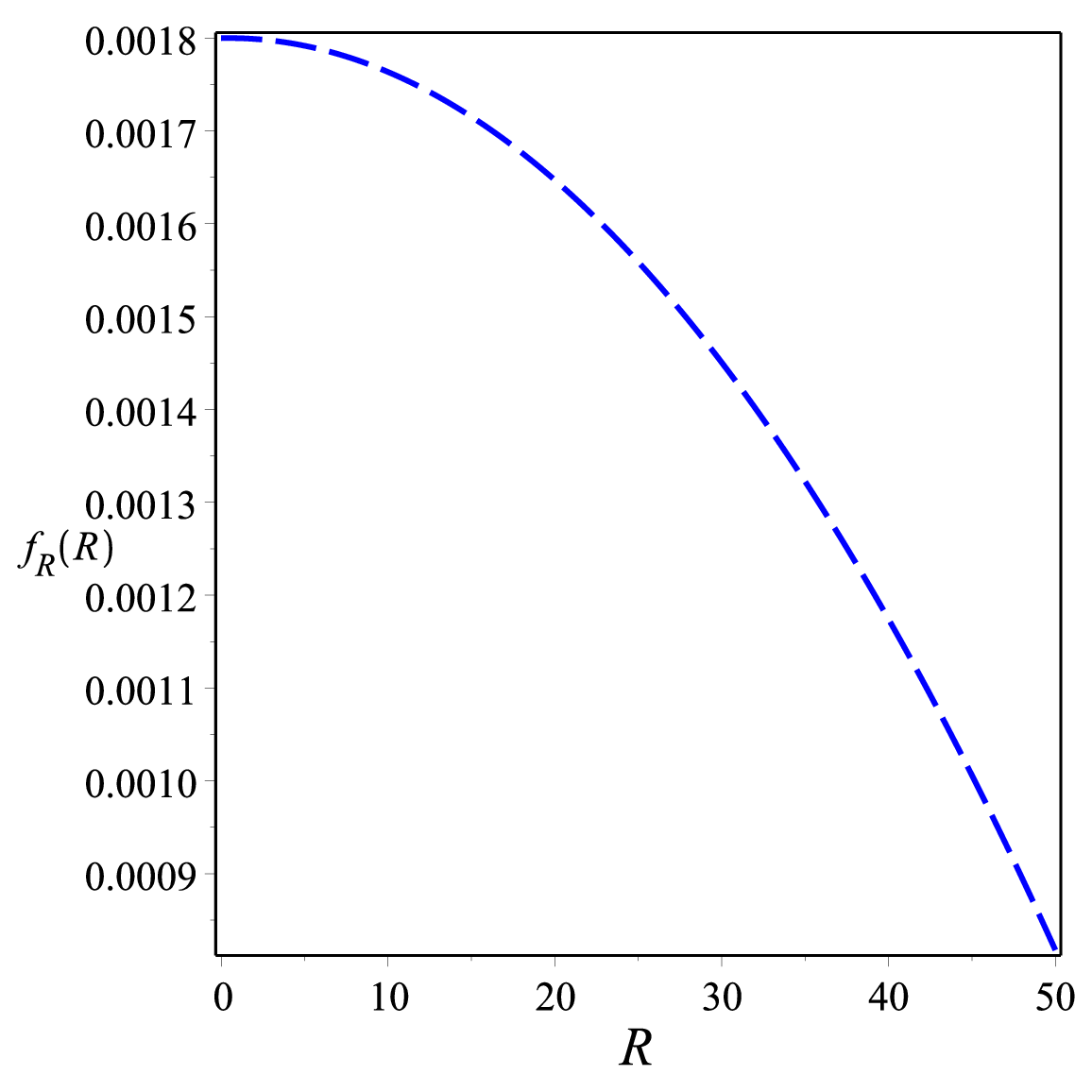}}
\subfigure[~ $f''(R)$ viz  $R$  ]{\label{fig:3c}\includegraphics[scale=0.25]{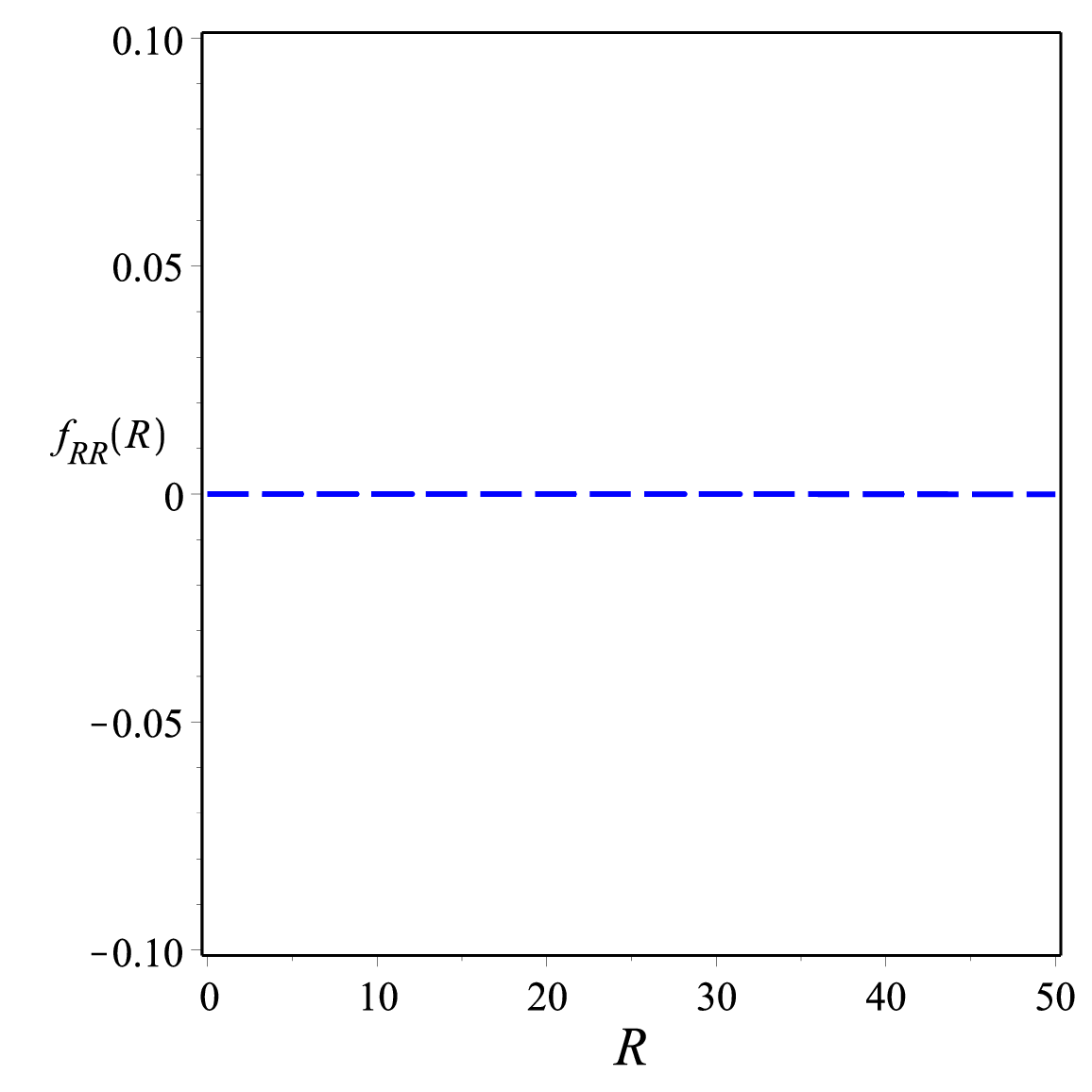}}
\caption{Schematic diagrams of  \subref{fig:3a} $f(R)$;  \subref{fig:3b} $f_R(R)$; \subref{fig:3c} $f_{RR}(R)$ viz $R$. We used the following numerical values to reproduce the above diagrams: $M=10$, $g=100$, $c_2=-0.01$ and $\alpha=10^2$}
\label{Fig:3}
\end{figure*}

Using  equation   ({ 8}) listed in the notebook  in Eq. (\ref{fa}) we get the form of $f(R)$ of the metric (\ref{mete}) as:
\begin{align}\label{f2R}
&f(R)=\frac{8769333362688c_1}{{g}^{33}{\alpha}^{9}} \left[  \left( {\frac {729}{478}}+{
\frac {729}{956}}\alpha{g}^{3}+{g}^{6}{\alpha}^{2} \right) {M}^{9}-{\frac {{g}^{33}{\alpha}^{8}}{
487185186816}}+{\frac {5{g}^{24}{\alpha}^{6
}{M}^{3}}{211451904}}-{\frac {31 {\alpha}^{3}{g}^{12}{M}^{6} }{183552}}\, \left( {\frac {54}{155}}+\alpha{g}^{3}
 \right)\right] R\nonumber\\
 &+\frac{8769333362688c_1}{{g}^{33}{\alpha
}^{9}}
 \left( {\frac {31}{1468416}}\,{\alpha}^{3}{g}^{15} \left( \alpha{g}^{3}+{\frac
{9}{31}} \right) {M}^{5}-3/16\, \left( {g}^{6}{\alpha}^{2}+{\frac {324}{239
}}+{\frac {162}{239}}\alpha{g}^{3} \right) {g}^{3}{M}^{8}-{\frac {5}{
3383230464}}{g}^{27}{\alpha}^{6}{M}^{2} \right) {R}^{2}\nonumber\\
 &+\frac{8769333362688c_1}{{g}^{33}{\alpha}^{9}} \bigg[\frac{1}{48} \left( {\frac {567}{956}}\alpha
{g}^{3}+{\frac {567}{478}}+{g}^{6}{\alpha}^{2} \right) {g}^{6}{M}^{7}+{
\frac {5{g}^{30}{\alpha\alpha}^{6}M}{121796296704}}-{\frac {155}{105725952}}
 \left( {\frac {36}{155}}+\alpha{g}^{3} \right){\alpha}^{3}{g}^{18}{M}^{4}
 \bigg] {R}^{3}\,.
 \end{align}
  It is easy to find the first and second derivatives of $f(R)$ w.r.t $R$ using Eq. (\ref{f2R}) to show if the model under consideration is stable or not. We show this behavior in Fig. \ref{Fig:3}. As the plots of  Fig. \ref{Fig:3} show that the generalized solution of Hayward in the frame of $f(R)$ is stable because $f(R)>0$, $f_R(R)>0$, and $f_{RR}(R)>0$. In the next section, we are going to study the patterns of the solution (\ref{sol22}).

 \section{Geodesic}\label{SeC.V}
In this section, our objective is to obtain the geodesic equation for the black hole described by Eq.~(\ref{sol22}).
Due to the spherical symmetry of the enhanced Hayward black hole, Our focus can be directed towards analyzing the movement of test particles within the equatorial plane, particularly focusing on $\theta = \frac{\pi}{2}$. Consequently, we have three geodesic equations to consider:
\newpage
\begin{equation}\label{7}
\frac{d^2t}{d\sigma^2}+\frac{h'(r)}{h(r)}\frac{dt}{d\sigma}\frac{dr}{d\sigma}=0\,,
\end{equation}

\begin{equation}\label{8}
\frac{d^2r}{d\sigma^2}+\frac{1}{2}h_1(r)h'(r)\Big(\frac{dt}{d\sigma}\Big)^2-\frac{[h(r)h_1(r)]'}{2h(r)h_1(r)}\Big(\frac{dr}{d\sigma}\Big)^2-
rh(r)h_1(r)\Big(\frac{d\phi}{d\sigma}\Big)^2=0\,,
\end{equation}

\begin{equation}\label{9}
\frac{d^2\phi}{d\sigma^2}+\frac{2}{r}\frac{d\phi}{d\sigma}\frac{dr}{d\sigma}=0\,,
\end{equation}
where $h(r)$ and $h_1(r)$ are defined in Eq.~(\ref{sol22}) and $\sigma$ is the affine parameter.
\\
In the following subsections, we will employ two methods to solve these geodesic equations. The first method involves utilizing the effective Newtonian potential, as explored in previous studies such as \cite{Cruz:2004ts, Abdujabbarov:2009az}. The second method involves employing dynamical system analysis, as described in \cite{Dahia:2007vd}.
\subsection{The efficient method of Newtonian potential}\label{sec4}
Lagrangian description the motion  in the generalized solution of the Hayward black hole is provided as:
\begin{align}\label{10}
&2\mathcal{L}=-\left(1-\frac{2Mr^2}{r^3+g^3}\right) \dot t^2+\frac{\dot r^2}{\left(1-\frac{2Mr^2}{r^3+g^3}\right)e^{\alpha r^3}}+r^2(\dot \theta^2+\sin^2\theta\dot\phi^2)\,.
\end{align}
The dot above indicates differentiation concerning the affine parameter $\sigma$. In this scenario,  $\mathcal{L}$ does not rely on  $t$ and $\phi$. Consequently, we derive the following two conserved quantities: the momentum $L$  and the energy $E$, which corresponds to $\phi$.\\
\vspace*{0.5cm}
\textbf{Energy}:
\begin{equation}\label{11}
E=g_{tt}\frac{dt}{d\sigma}=h(r)\frac{dt}{d\sigma}\,.
\end{equation}
\textbf{Momentum}:
\begin{equation}\label{12}
2L=\frac{\partial{\mathcal{L}}}{\partial{\dot\phi}}=2r^2\dot\phi=constant\,.
\end{equation}
In this context, $L$ denotes the angular momentum. Taking into account the orthogonality condition, or keeping in mind the need for appropriate normalization,
\begin{equation}\label{13}
g_{\mu\nu}\frac{dx^{\mu}}{d\sigma}\frac{dx^{\nu}}{d\sigma}=-\varepsilon\,.
\end{equation}
 In the given context, the parameter $\varepsilon$ assumes a value of $\varepsilon = 1$ for timelike geodesics and $\varepsilon = 0$ for null geodesics (as stated in  Ref. \cite{Dahia:2007vd}). Considering the equatorial plane, we have the following
\begin{equation}\label{144}
\Big(\frac{dr}{d\sigma}\Big)^2=E^2h_1(r)-h(r)h_1(r)\Big(\frac{L^2}{r^2}+\varepsilon\Big)\,.
\end{equation}
Hence, equations (\ref{11}), (\ref{12}), and (\ref{144}) represent the necessary equations for characterizing the movement of particles along the equatorial plane of the extended regular BH.

Next, we will proceed to rewrite Eq. (\ref{144}) as follows:
\begin{align}\label{15}
&\frac{1}{2}\Big(\frac{dr}{d\sigma}\Big)^2=E_{eff}-V_{eff}\,, \quad \mbox{where} \quad
E_{eff}=\frac{E^2}{2}\,\quad \mbox{and}\nonumber\\
&V_{eff}=hh_1\Big(\frac{L^2}{r^2}+\epsilon\Big)=\left(1-\frac{2Mr^2}{r^3+g^3}\right)e^{\alpha r^3}\Big(\frac{L^2}{r^2}+\epsilon\Big)\nonumber\\
&\approx  \frac{L^2}{r^2}+\left(\varepsilon-\frac{2ML^2}{g^3}\right)+\alpha L^2 r-\frac{2M}{g^3}r^2+\left[\varepsilon\alpha-\frac{2M\left(\alpha g^3-1\right)L^2}{g^6}\right]r^3
\,.
\end{align}

When we introduce the functions $h(r)$ and $h_1(r)$, Eq.~(\ref{144}) can be expressed in the following manner:
\begin{align}\label{18}
&\Big(\frac{dr}{d\sigma}\Big)^2\approx E^2-\frac{L^2}{r^2}-\left(\varepsilon-\frac{2ML^2}{g^3}\right)-\alpha L^2 r+\frac{2M}{g^3}r^2-\left[\varepsilon\alpha-\frac{2M\left(\alpha g^3-1\right)L^2}{g^6}\right]r^3\,.
\end{align}
The configuration of geodesics within the equatorial plane ($\theta = \frac{\pi}{2}$) can be established
  as:
\begin{equation}\label{19}
\frac{dr}{d\sigma}=\frac{dr}{d\phi}\frac{d\phi}{d\sigma}=\frac{L}{r^2}\frac{dr}{d\phi}\,.
\end{equation}
Next, we introduce the transformation $u = \frac{1}{r}$, resulting in:
\begin{equation}\label{20}
\Big(\frac{du}{d\sigma }\Big)^2=u^4\Big(\frac{dr}{d\sigma}\Big)^2\,.
\end{equation}
By substituting Eq.~(\ref{20}) into Eq.~(\ref{18}), we obtain:
\begin{align}\label{21}
\Big(\frac{du}{d\sigma}\Big)^2=& \left[E^2-\left(\varepsilon{+}\frac{2ML^2}{g^3}\right)\right]u^4-L^2 u^6-\alpha L^2 u^3+\frac{2M}{g^3}u^2-\left[\varepsilon\alpha-\frac{2M\left(\alpha g^3-1\right)L^2}{g^6}\right]u\,.
\end{align}
By utilizing Eqs.~(\ref{19}), (\ref{20}), and (\ref{21}), we derive the expression for $\Big(\frac{du}{d\phi}\Big)^2$ in the following manner:
\begin{align}\label{22}
\Big(\frac{du}{d\phi}\Big)^2=&\frac{E^2-\left(\varepsilon+\frac{2ML^2}{g^3}\right)}{L^2}- u^2-\frac{\alpha }{u}+\frac{2M}{g^3u^2L^2}-\frac{\varepsilon\alpha-\frac{2M\left(\alpha g^3-1\right)L^2}{g^6}}{u^3L^2}=S(u)\,.
\end{align}
Equation (\ref{22}) provides a description of the trajectories followed by test particles in the vicinity of the extended regular black hole. From a physical perspective, analyzing the trajectories of particles involves investigating two essential aspects: their radial motion and their circular motion. Therefore, in the upcoming section, we will examine the trajectories of particles in these two scenarios.
\subsection{Radial motion}\label{sec5}
Analyzing the radial geodesics is crucial for understanding the fundamental characteristics of the spacetime, particularly when the angular momentum is zero. In the case of radial motion ($\phi = \text{constant}$), Eq.~(\ref{22}) does not provide useful information about the radial trajectories, as certain terms tend toward infinity $L\to 0$. To overcome we rewrite Eq. (\ref{144}) as:
\begin{equation}\label{14}
\Big(\frac{dr}{d\sigma}\Big)^2=E^2h_1(r)-h(r)h_1(r)\epsilon\,.
\end{equation}
\subsubsection{Movement of massive particles}\label{sec6}
When considering massive particles with $\epsilon = 1$, Eq. (\ref{14}) results in:
\begin{align}\label{24}
&\Big(\frac{dr}{d\sigma}\Big)^2\approx (E^2-1)(1+\alpha r^3)+\frac{2M}{g^3}r^2\,.
\end{align}
By taking the derivative of Eq. (\ref{24}) with respect to $\sigma$, we acquire:
\begin{equation}\label{25}
\frac{d^2r}{d\sigma^2}= \frac{3}{2}(E^2-1)\alpha r^2+\frac{2M}{g^3}r\,.
\end{equation}
We use the proper time parameter $\tau$ instead of $\sigma$ along the trajectory of massive particles because their paths follow timelike geodesics. This lets us use the previously mentioned equation to investigate the motion of massive particles. Furthermore, the following is the criterion for an attractive force per unit mass:
\begin{equation}\label{26}
 \alpha r>\frac{4M}{3g^3(1-E^2)} \,, \quad \mbox{provided} \, \,\alpha>0 \,, \quad \alpha r>\frac{4M}{g^3(E^2-1)} \,, \quad \mbox{provided} \, \,\alpha<0 \,.
\end{equation}
The condition provided by Eq. (\ref{26}) holds significance as it determines the existence of bound states for massive particles. When a particle passes through the gravitational field of a black hole, it can acquire kinetic energy through gravitational interactions. When calculating the variation in potential energy encountered by the particle within the gravitational field, we evaluate this change from the particle's resting position, where the radial coordinate $r$ transitions to $R$.
Our attention is now directed towards depicting the particle's trajectories in the $(t,r)$-plane. To accomplish this, we can make use of Equations (\ref{11}) and (\ref{24}).
\begin{equation}\label{28}
\Big(\frac{dr}{dt}\Big)^2=\Big[E^2h_1(r)-h(r)h_1(r)\Big]\frac{h^2(r)}{E^2}\,.
\end{equation}
The condition $\frac{dr}{dt} = 0$ holds for all points when $E^2 = h(r)$. Utilizing Equation (\ref{24}), we can deduce the effective potential $V_{\text{eff}}$ as follows:
\begin{equation}\label{29}
V_{eff}=\frac{1}{2}\Bigg(1+\alpha r^2-\frac{3}{2}E^2\alpha r^2-\frac{2M}{g^3}r\Bigg)\,.
\end{equation}
\begin{figure}[ht]
\centering
\subfigure[~Potential given by Eq. (\ref{29}) of the BH (\ref{sol22}). Here we put $M=10$, $E=0.1$ and $g=10$.]{\label{fig:4a}
\includegraphics[scale=0.25]{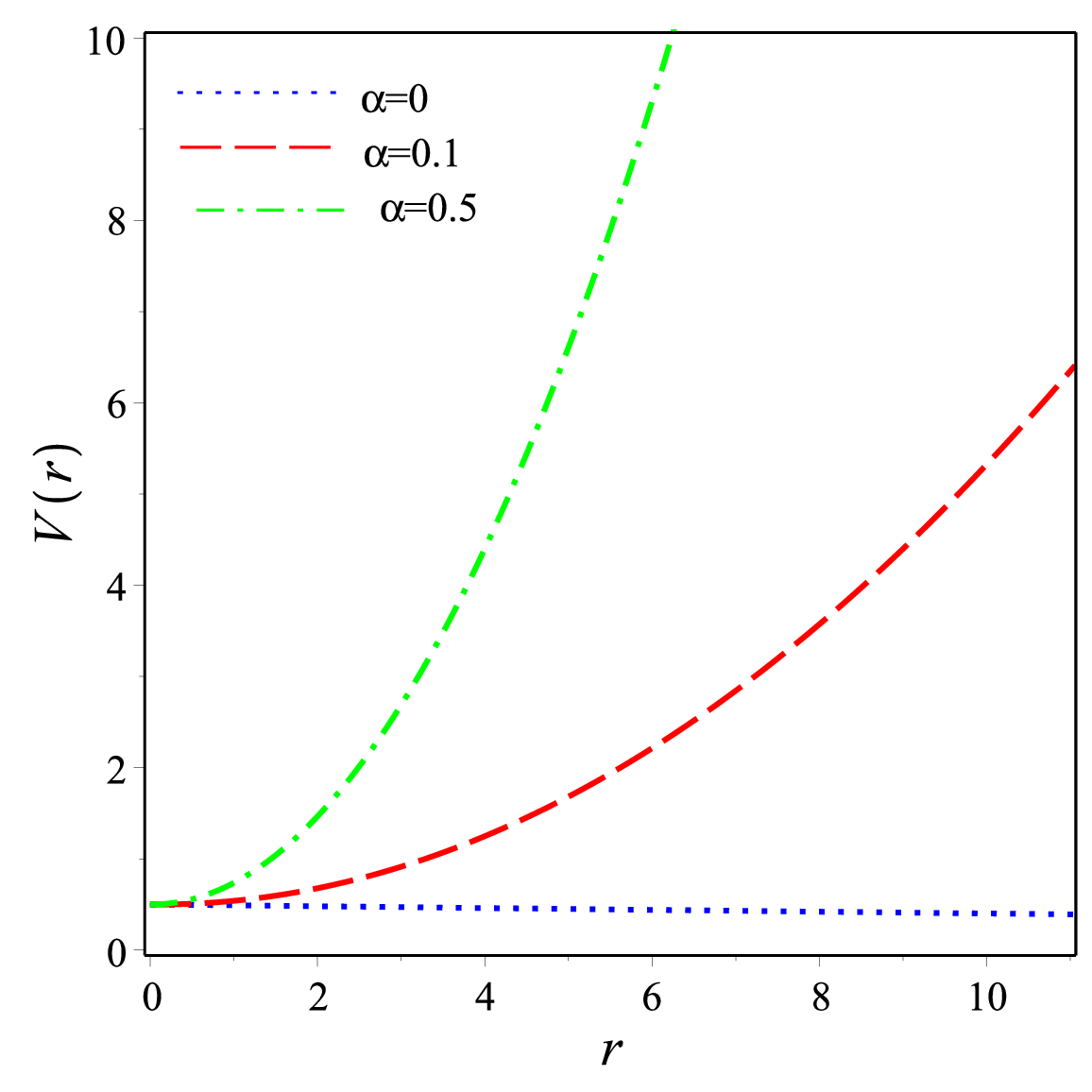}}\hspace{0.3cm}
\caption{ {represents the potential given by Eq.~(\ref{29}) as $g\to 0$ for the black hole (\ref{sol22}. The plots are shown in relation to the BH's coordinate $r$ (\ref{sol22}).}}
\label{Fig:4}
\end{figure}
Figure \ref{Fig:4} demonstrates how the effective potential $V_{\text{eff}}$ changes with the radius $r$ for three different values of the dimensional parameter $\alpha$.
From Eq. (\ref{28}), we can express it as follows:
\begin{equation}\label{30}
\frac{dr}{dt}=\frac{h(r)}{E}\sqrt{E^2h_1(r)-h(r)h_1(r)}\,.
\end{equation}
In the case when $\alpha = 0$, Eq.~(\ref{28}) does not become singular as $r \rightarrow 0$.
\subsubsection{The movement of photons}\label{sec7}
When dealing with the motion of photons with $\epsilon = 0$, Eq. (\ref{14}) yields:
\begin{equation}\label{31}
\Big(\frac{dr}{d\epsilon}\Big)^2=E^2h_1(r)\,,
\end{equation}
where, $E$ is defined in Eq.~(\ref{11}). Additionally, by defining $\frac{dr}{dt}=\frac{dr}{d\sigma}\frac{d\sigma}{dt}$, we obtain:
\begin{align}\label{32}
&\frac{dr}{dt}=\pm E\big\{1+\alpha r^3+\frac{1}{2}\alpha^2r^6\big\}\,.
\end{align}
The solution of Eq.~(\ref{32}) is complex, making it challenging to determine the trajectories of photons using this approach.\\
{\bf Motion in a circular path}\label{sec8}:\\
Equation (\ref{22}) demonstrates that $u$ is constant at the equilibrium position of circular orbits. It is thus possible to deduce $S(u) = 0$ and $S'(u) = 0$. Using Equation (\ref{22}), we arrive at:
\begin{align}\label{33}
S(u)=&\frac{E^2-\left(\varepsilon+\frac{2ML^2}{g^3}\right)}{L^2}- u^2-\frac{\alpha }{u}+\frac{2M}{g^3u^2L^2}-\frac{\varepsilon\alpha-\frac{2M\left(\alpha g^3-1\right)L^2}{g^6}}{u^3L^2}\,,
\end{align}
and the derivative of Eq. (\ref{33}) yields:
\begin{align}\label{34}
S'(u)=&- 2u+\frac{\alpha }{u^2}+\frac{4M}{g^3u^3L^2}+3\frac{\varepsilon\alpha-\frac{2M\left(\alpha g^3-1\right)L^2}{g^6}}{u^4L^2}\,.
\end{align}
Upon solving Equations (\ref{33}) and (\ref{34}), we obtain the following expressions for the energy ($E$) and angular momentum ($L$) of a particle moving in a circle:
\begin{eqnarray}\label{35}
L^2={\frac {{r}^{4}( {4\,M-3\,{g}^{3}\alpha\,r})}{{{g}^{3}\alpha {
r}^{3}-2\,{g}^{3}-6\,M\alpha{g}^{3}{r}^{5}+6\,M{r}^{5}}}}\,,
\end{eqnarray}
and
\begin{align}\label{36}
&E^2=2\frac {{4\,{M}^{2}{r}^{4}-{g}^{6}+4\,M{g}^{3}{r}^{2}-
2\alpha{g}^{6}{r}^{3}-3\,M\alpha{g}^{6}{r}^{5}-2\,M\alpha{g}^{3}{r}^{5}-{\alpha}^{2}{g}^
{6}{r}^{6}+3\,M{g}^{3}{r}^{5}-2\,{M}^{2}{r}^{7}+2\,{M}^{2}\alpha{g}^{3}{r}^
{7}}}{{g}^{3}( {{g}^{3}\alpha{r}^{3}-6\,M\alpha{g}^{3}{r}^{5}+6
\,M{r}^{5}-2\,{g}^{3}})}\,.
\end{align}

\section{Conclusion and discussion}
In summary, investigating the Hayward solution of regular black holes in $f(R)$ gravity with a scalar field has revealed many interesting things about black hole physics. The Hayward solution is a revision to the traditional Schwarzschild solution that includes non-singular behavior at the black hole's center.

The potential of $f(R)$ gravity to resolve some of the central issues and questions raised by general relativity, including the nature of singularities and the existence of black holes, is one of the main reasons to consider it. The $f(R)$ gravity with a scalar field $\chi$ provides a promising avenue with an extra degree for investigating alternative theories of gravity by introducing modifications to the Einstein-Hilbert action through a function of the Ricci scalar.

Interesting properties set the Hayward solution apart from the conventional solutions in the frame of $f(R)$ gravity alongside with scalar field. Following the application of the $f(R)$ field equation with scalar field to a spherically symmetric spacetime, we separated the corresponding system of non-linear differential equations into two cases:
\begin{itemize}
\item[{\textrm a-}] The situation where the metric concedes with the Hayward metric when there is equal metric ansatz. In this instance, we extract the function of the scalar field $\psi$, the corresponding forms of $f(R)$, and the scalar field $\chi$.What's remarkable in this case is that $f(R)$ doesn't have a cosmological term, unlike what we've seen in typical black holes before \cite{Nashed:2022xmv,Nashed:2021hgn}. This indicates that we have the case of GR and the Schwarzschild solution if the Ricci scalar $R$ equals zero. Equation (\ref{ricci1}) can be used to realize this case, as $g=0$ yields $R=0$.\\
    \item[{\textrm b-}] We derive a black hole different from the Hayward one in the case of unequal metric potentials. We demonstrate that this black hole solution is an extension of the Hayward black hole solution and depends on a dimensional quantity $\alpha$. We obtained the standard Hayward black hole when $\alpha=0$. The corresponding forms of $f(R)$, $\chi$, and $\psi$ are also derived. We furthermore, demonstrate that the form of $f(R)$ was independent of any cosmological constant.
\end{itemize}

For each case, we compute the metric's form as $r \to 0$ and demonstrate that the metric is finite as $r\to 0$. Additionally, we computed the invariants for each case and demonstrated that they are all finite as $r\to 0$. Additionally, we demonstrated via plots that the first and second derivatives of $f(R)$ have a positive form.   Furthermore, we demonstrated that the Hayward solution extension offers insights into the implications of modified gravity on the event horizon.

Next, we examine the paths taken by these solutions and determine the potential form for both scenarios. We demonstrated that the potential falls with an increase in $\alpha$ in the extension Hayward case. The form of these solutions' conserved quantities, $E$ and $L$, was also deduced.

 We can learn more about potential changes to Einstein's theory and how it might be reconciled with other fundamental theories like quantum mechanics by investigating alternative theories of gravity, such as $f(R)$.

In conclusion, studying the Hayward solution a regular black hole solution in the context of $f(R)$ gravity combined with a scalar field offers important new understandings of the characteristics and behavior of black holes in alternative theories of gravity. In this study, we assume the form of the metric potential which describes a regular black hole, and derive its corresponding $f(R)$. Can we assume a specific form of $f(R)$ that from it we can derive the metric potentials that describe a regular black hole? This study can be carried out elsewhere.


%

\end{document}


\title{Extension of Hayward black hole in   $f(R)$ gravity coupled with a scalar field : Supplementary material}

\author{G.~G.~L.~Nashed~\orcid{0000-0001-5544-1119}}
\email{nashed@bue.edu.eg}
\affiliation {Centre for Theoretical Physics, The British University in Egypt, P.O. Box
43, El Sherouk City, Cairo 11837, Egypt}
\date{\today}

\begin{abstract}
 We report  here, for the interested reader,  the various solution components. In particular, the two functions $F(r)$, $\psi(r)$, the invariants,  $\xi$, $f(R)$ the function $f(r)$, the Ricci scalar $R$.
\end{abstract}
\maketitle

\section{The explicate form of the scalar function when $h_1=1$ }

\begin{align}\label{sca1}
&126\, \bigg\{  \left( -{\frac {5}{21}}\,{r}^{9}+4/7\,M{r}^{8}-2/7\,{r}^ {6}{g}^{3}+M{r}^{5}{g}^{3}+1/7\,{r}^{3}{g}^{6}+{\frac {4}{21}}\,{g}^{9 } \right)  \left( -1+\chi' \right)  \left( r+g \right)  \left( -1/3\,{g}^{6}-2/3\,{g}^{3}{r}^{3} -1/3\,{r}^{6}+{r}^{5}M \right) \nonumber\\
 &\left(M{r} ^{2}  -1/2{g}^{3}-1/2{r}^{3}\right) r \left( {r}^{2}-rg+{g}^{2} \right) \chi' +4 \chi'    \bigg[ \left( r+g \right)  \bigg( {\frac {5}{84}}\,{r}^{15}-{\frac {5}{84}}\,{g}^{12}{r}^{3}-{\frac {11}{42}}\, {g}^{9}{r}^{6}-{\frac {4}{21}}\,{g}^{6}{r}^{9}+{\frac {5}{14}}{r}^{ 13}{M}^{2}-2/7{r}^{14}M\nonumber\\
 &+{\frac {5}{14}}\,{g}^{6}{r}^{8}M-4/7\,{g}^{3 }{r}^{11}M+1/42\,{g}^{15}+{g}^{3}{r}^{10}{M}^{2}+1/21\,{g}^{3}{r}^{12} +{\frac {9}{14}}\,{g}^{9}{r}^{5}M \bigg)  \left( -1/2\,{g}^{3}-1/2\,{ r}^{3}+M{r}^{2} \right)  \left( {r}^{2}-rg+{g}^{2} \right) \chi' \nonumber\\
 &+{\frac {5}{126}}{r}^{21}-{\frac {5} {18}}\,{r}^{20}M+{\frac {19}{28}}{M}^{2}{r}^{19}+ \left( {\frac {4}{ 63}}{g}^{3}-\frac{4M^{3}}7 \right) {r}^{18}-3/4{g}^{3}{r}^{17}M+{ \frac {31}{14}}{g}^{3}{r}^{16}{M}^{2}-\left( {\frac {53}{28}}{g} ^{3}{M}^{3}+{\frac {5}{21}}{g}^{6} \right) {r}^{15}+{\frac {13}{42}} {g}^{6}{r}^{14}M\nonumber\\
 &+{\frac {27}{28}}{g}^{6}{r}^{13}{M}^{2}- \left( { g}^{6}{M}^{3}+{\frac {95}{126}}{g}^{9} \right) {r}^{12}+{\frac {160} {63}}{g}^{9}{r}^{11}M-2{g}^{9}{r}^{10}{M}^{2}-{\frac {95}{126}}\,{ g}^{12}{r}^{9}+{\frac {17}{7}}{g}^{12}{r}^{8}M-{\frac {10}{7}}{g}^ {12}{r}^{7}{M}^{2}-{\frac {5}{21}}{g}^{15}{r}^{6}\nonumber\\
 &+{\frac {47}{84}}\, {g}^{15}{r}^{5}M+{\frac {4}{63}}\,{g}^{18}{r}^{3}-1/9\,{g}^{18}{r}^{2} M+{\frac {5}{126}}\,{g}^{21} \bigg]  \bigg\} c_1\exp\bigg\{-2 \,\int \! \bigg\{ 63\, \bigg[4/7M{r}^{8} -{\frac {5}{21}}{r}^{9}-2 /7{r}^{6}{g}^{3}+M{r}^{5}{g}^{3}+1/7{r}^{3}{g}^{6}\nonumber\\
 &+{\frac {4}{21}} {g}^{9} \bigg] \left({r}^{5}M-\frac{1}3{g}^{6}-\frac{2}3{g}^{3}{r}^{3}-\frac{1}3{r}^{ 6} \right) r\chi'' +252 \bigg[ {\frac {5}{84}}{r}^{15}-{\frac {5}{84}}{g}^ {12}{r}^{3}-{\frac {11}{42}}{g}^{9}{r}^{6}-{\frac {4}{21}}{g}^{6}{ r}^{9}+{\frac {5}{14}}{r}^{13}{M}^{2}-\frac{2}7{r}^{14}M+{\frac {5}{14}} {g}^{6}{r}^{8}M\nonumber\\
 &-4/7{g}^{3}{r}^{11}M+1/42{g}^{15}+{g}^{3}{r}^{10} {M}^{2}+1/21{g}^{3}{r}^{12}+{\frac {9}{14}}{g}^{9}{r}^{5}M \bigg]\chi'  \bigg\} \chi'^{-1}{r}^{-1} \left( -{g}^{6}-2{g}^{3}{r}^{3}-{r}^{6}+3{r}^{5}M \right) ^{-1} \bigg[12M{r}^{8} -5{r}^{9}\nonumber\\
 &-6\,{r}^{6}{g}^{3}+21\,M{r}^{5}{g}^{3} +3\,{r}^{3}{g}^{6}+4\,{g}^{9} \bigg]^{-1}{dr}\bigg\}  \left( {r}^{3}+{g}^{ 3} \right) ^{-2} \left( -{g}^{6}-2\,{g}^{3}{r}^{3}-{r}^{6}+3\,{r}^{5}M \right) ^{-1} \bigg[ 12\,M{r}^{8} -5\,{r}^{9}-6\,{r}^{6}{g}^{3}\nonumber\\
 &-+21\,M {r}^{5}{g}^{3}+3\,{r}^{3}{g}^{6}+4\,{g}^{9} \bigg]^{-1}{r}^{-1}=0\,.
\end{align}
\section{The explicate forms of $F(r)$, $\psi(r)$ and $\chi(r)$ when $h_1\neq 1$}\label{Sec:App_1}
The exact solution of the first three differential Eqs.~({\color{blue}9}), presented in the manuscript,  when $h_1\neq 1$ takes the form:
\begin{align}\label{sol2}
&F(r)= 18\, \left( r+g \right)^{2} \left( {r}^{2}-gr+{g}^{2} \right)^{2} h_1^{2}{r}^{2}c_2\, \chi'^{2}\exp\bigg\{\int {\frac {1}{972}}\, \bigg[ -729\, \bigg[ 4/9\, \left( r +g \right)  \left( {r}^{2}-gr+{g}^{2} \right) r \bigg( -1/3\,\alpha\,{ r}^{12}+M{r}^{11}\alpha\nonumber\\
&+ \left( -2/9-\alpha\,{g}^{3} \right) {r}^{9}+M {r}^{8}\alpha\,{g}^{3}+ \left( -2/3\,{g}^{3}-\alpha\,{g}^{6} \right) { r}^{6}+4\,{g}^{3}{r}^{5}M+ \left( -2/3\,{g}^{6}-1/3\,\alpha\,{g}^{9} \right) {r}^{3}-2/9\,{g}^{9} \bigg) \bigg( -1/3\,{g}^{6}-2/3\,{g}^{ 3}{r}^{3}\nonumber\\
&+{r}^{5}M-1/3\,{r}^{6} \bigg)\chi'' + \chi' \bigg( 1/9\,{r}^{24}{\alpha}^{2}-2/3\,{r}^{23}M{ \alpha}^{2}+{r}^{22}{M}^{2}{\alpha}^{2}+ \left( 2/3\,{g}^{3}{\alpha}^{ 2}+{\frac {20}{243}}\,\alpha \right) {r}^{21}-8/3\, \left( \alpha\,{g} ^{3}+1/27 \right) \alpha\,M{r}^{20}\nonumber\\
&+2\,\alpha\, \left( \alpha\,{g}^{3} -2/9 \right) {M}^{2}{r}^{19}+ \left( {\frac {40}{81}}\,\alpha\,{g}^{3} +5/3\,{g}^{6}{\alpha}^{2}+{\frac {44}{729}} \right) {r}^{18}-4\,\alpha \, \left( \alpha\,{g}^{3}+{\frac {56}{81}} \right) {g}^{3}M{r}^{17}+ \alpha\, \left( \alpha\,{g}^{3}+{\frac {56}{9}} \right) {g}^{3}{M}^{2} {r}^{16}\nonumber\\
&+{\frac {20}{9}}\, \left( {g}^{6}{\alpha}^{2}+5/9\,\alpha\,{g} ^{3}+{\frac {22}{135}} \right) {g}^{3}{r}^{15}-8/3\,{g}^{3} \left( { \frac {26}{9}}\,\alpha\,{g}^{3}+{\frac {38}{27}}+{g}^{6}{\alpha}^{2} \right) M{r}^{14}+{\frac {20}{3}}\, \left( \alpha\,{g}^{3}+{\frac {32 }{45}} \right) {g}^{3}{M}^{2}{r}^{13}+5/3\, \left( {\frac {44}{81}}\right.\nonumber\\
&\left.+{g }^{6}{\alpha}^{2}+{\frac {80}{81}}\,\alpha\,{g}^{3} \right) {g}^{6}{r} ^{12}-2/3\, \left( {g}^{6}{\alpha}^{2}+{\frac {104}{9}}+{\frac {304}{ 27}}\,\alpha\,{g}^{3} \right) {g}^{6}M{r}^{11}+{\frac {368}{27}}\,{g}^ {6}{r}^{10}{M}^{2}+2/3\, \left( {g}^{6}{\alpha}^{2}+{\frac {50}{27}}\, \alpha\,{g}^{3}+{\frac {440}{243}} \right) {g}^{9}{r}^{9}\nonumber\\
&-{\frac {200} {81}}\,{g}^{9} \left( \alpha\,{g}^{3}+{\frac {42}{25}} \right) M{r}^{8 }+1/9\, \left( {g}^{6}{\alpha}^{2}+{\frac {220}{27}}+{\frac {40}{9}}\, \alpha\,{g}^{3} \right) {g}^{12}{r}^{6}-{\frac {16}{81}}\,{g}^{12}{r}^ {5}M+ \left( {\frac {88}{243}}\,{g}^{15}+{\frac {20}{243}}\,{g}^{18} \alpha \right) {r}^{3}+{\frac {44}{729}}\,{g}^{18} \bigg)  \bigg]\nonumber\\
&\left( r+g \right) ^{2} \left( {r}^{2}-gr+{g}^{2} \right) ^{2}h_1^2+ \bigg\{-1944\, \bigg( {\frac {1}{54}}\, {r}^{21}\alpha-1/18\,M\alpha\,{r}^{20}+ \left( {\frac {2}{81}}+1/9\, \alpha\,{g}^{3} \right) {r}^{18}-5/9\,{g}^{3}{r}^{17}M\alpha+{g}^{3}{r }^{16}{M}^{2}\alpha\nonumber\\
&+ \left( {\frac {5}{18}}\,\alpha{g}^{6}+{\frac {4 }{27}}{g}^{3} \right) {r}^{15}-4/3{g}^{3} \left( 4/3+\alpha{g}^{ 3} \right) M{r}^{14}+ \left( \alpha{g}^{3}+8/3 \right) {g}^{3}{M}^{2 }{r}^{13}+{\frac {10}{27}}{g}^{6} \left( \alpha\,{g}^{3}+1 \right) { r}^{12}-{\frac {11}{9}} \left( {\frac {30}{11}}+\alpha{g}^{3} \right) {g}^{6}M{r}^{11}\nonumber
\end{align}
\newpage
\begin{align}
&+14/3\,{g}^{6}{r}^{10}{M}^{2}+ \left( {\frac {5}{18}}\,{g}^{12}\alpha+{\frac {40}{81}}\,{g}^{9} \right) {r}^{9}-{ \frac {7}{18}}\, \left( \alpha\,{g}^{3}+{\frac {24}{7}} \right) {g}^{9 }M{r}^{8}+ \left( {\frac {10}{27}}\,{g}^{12}+1/9\,{g}^{15}\alpha \right) {r}^{6}+\frac{2}9\,{g}^{12}{r}^{5}M+{\frac {2}{81}}\,{g}^{18}\nonumber\\
&+{\frac {1}{54}}{g}^{15} \left( 8+\alpha{g}^{3} \right) {r}^{3} \bigg) r \left( -1/3{g}^{6}-2/3{g}^{3}{r}^{3}+{r}^{5}M-1/3{r}^{6} \right)\chi'' -1458 \, \chi' \bigg[ -{\frac {4}{243}}\,{g}^{24}+{\frac {224}{9}} \bigg( -{\frac {5}{216} }\alpha\,{g}^{6}-{\frac {1}{72}}{g}^{9}{\alpha}^{2}+ {M}^{3}\nonumber\\
& -\frac{1}{27}{g}^{3}\bigg) {g}^{6}{r}^{15}-{\frac {14}{81}}\left( {g}^{6}{ \alpha}^{2}+{\frac {20}{3}}+\frac{8}3\alpha\,{g}^{3} \right) {g}^{12}{r}^{ 12}-{\frac {4}{81}} \left( {g}^{6}{\alpha}^{2}+\frac{14}3\alpha{g}^{3} +{\frac {56}{3}} \right) {g}^{15}{r}^{9}-{\frac {1}{162}} \left( {g} ^{6}{\alpha}^{2}+{\frac {32}{3}}\alpha{g}^{3}+{\frac {224}{3}} \right) {g}^{18}{r}^{6}\nonumber\\
&-{\frac {2g^{21}r^3 }{243}}\left( 16+\alpha g^3 \right) -\frac{1}{18}r^{28}M^2\alpha^2+{\frac {248} {27}}g^{12}r^{10}{M}^{2}+\frac{1}{27}{r}^{29}M \alpha ^2+ \bigg[  \left(M^3\alpha^2-\frac {56}{243}\alpha \right) g^3-{ \frac {28}{81}}g^6\alpha^2-{\frac {32}{243}}+4\alpha M^{3} \bigg] g^3r^{21}\nonumber\\
&+8\, \left({\frac {10}{9}}\,{M}^{3}  -{\frac {35}{648}}{g}^{ 9}{\alpha}^{2}-{\frac {14}{243}}\alpha g^6+ \left(\alpha M^3 -{\frac {14} {243}} \right) g^3\right) g^3r^{18}+{\frac {32}{81}}g^{18}r^5M-{\frac {1} {162}}\,{r}^{30}{\alpha}^{2}+ r^{24}\bigg[ {g}^{3}\left( {M}^{3}{\alpha}^{2}-{\frac {16}{243}}\alpha \right)-{\frac {4}{243}}\nonumber\\
& -{\frac {14}{81} }g^6\alpha^2 \bigg] -\left( {\frac {4}{81}}g^3\alpha^2+{\frac {2}{243}}\alpha \right) r^{27}+\frac{g^3 r^{26}M\alpha^2}3-\frac {13 g^6M^2r^{16}}{18} \left( {\frac {464}{13}}+{ \frac {32}{3}}\alpha g^3+g^6\alpha^2 \right)+{\frac {7 g^9M r^{14}}{9}}\bigg( {\frac {320}{63}}+g^6 \alpha^2\nonumber\\
&+{\frac {64}{21}}\alpha g^3 \bigg)-{ \frac {46}{27}} \left( {\frac {96}{23}}+\alpha g^3 \right) g^9M^2r^{13}+{\frac {4}{27}}{g}^{12}M \left( {g}^{6}{\alpha}^{2}-{\frac {28}{3}}-\alpha\,{g}^{3} \right) {r}^{11}-{\frac {8}{27}}\, \left( \alpha\,{g}^{3}+7/3 \right) {g}^{15}M{r}^{8}-{\frac {20}{9}}\, {g}^{3} \bigg( {g}^{6}{\alpha}^{2}+{\frac {64}{15}}\nonumber\\
&+{\frac {23}{5}} \alpha g^3\bigg) M^2 r^{19}+5/3 \left( {\frac {136}{45}} \alpha g^3+g^6\alpha^2+{\frac {592}{135}} \right) {g}^{ 6}M{r}^{17}-{\frac {8}{9}}\, \left( \alpha\,{g}^{3}-1/12 \right) \alpha\,{M}^{2}{r}^{25}+{\frac {50}{27}}\,{g}^{3} \bigg( {\frac {124}{ 75}}+{g}^{6}{\alpha}^{2}+{\frac {52}{25}}\alpha g^3 \bigg) Mr^{20}\nonumber\\
&+{\frac {10}{9}}\,\alpha\,{g}^{3}M \left( \alpha\,{g}^{3}+{ \frac {14}{15}} \right) {r}^{23}-7/3\, \left( \alpha\,{g}^{3}+{\frac { 16}{9}} \right) \alpha\,{g}^{3}{M}^{2}{r}^{22} \bigg]  \bigg\} h_1^3-144\, \left( r+g \right) ^{5} \left( {r} ^{2}-gr+{g}^{2} \right) ^{5}\bigg\{  \bigg[  \bigg( 1/6\,{g}^{3}{r}^{6}M-1/18\,{g}^{9}r \nonumber\\
&-1/ 6r^4g^6-1/6r^7g^3-1/18r^{ 10}+1/6r^9M \bigg)\chi'' + \chi'\bigg( -1/3\alpha r^{12}+Mr^{11}\alpha+ \left( -\alpha g^3-1/6 \right) r^9+Mr^8\alpha g^3- \left( \alpha{g}^{6} +1/2g^3 \right) {r}^{6}\nonumber\\
&+3g^3r^5M-\left( \frac{1}3\alpha g^9+1/2g^6 \right) r^3-\frac{1}6g^9 \bigg)  \bigg] h_1+\frac{1}{36}\chi' \left( r+g \right)^3 \left( r^2-gr+ g^2\right) ^3 \bigg\}  \bigg] \chi'^{-1} \left( r^5M-\frac{1}3g^6-2/3g^3r^3-\frac{1}3r^6 \right) ^{-1}{r}^{-1}\nonumber\\
&\bigg[ 1/6 \left(M{r}^{11}\alpha - 1/3\alpha{r}^{12}+ \left( -2/9-\alpha {g}^{3} \right) {r}^{9}+M{r}^{8}\alpha{g}^{3}-\left( 2/3{g}^{3}+\alpha {g}^{6} \right) {r}^{6}+4{g}^{3}{r}^{5}M-\left(2/3{g}^{6}+1/3 \alpha{g}^{9} \right) {r}^{3}-2/9{g}^{9} \right)\nonumber\\
&  \left( r+g \right)^3 \left( r^2-gr+g^2 \right) ^3h_1^2+ \bigg[ {\frac {1}{54}}r^{21}\alpha-1/18M \alpha r^{20}+ \left( {\frac {2}{81}}+\frac{1}9\alpha g^3 \right) r^{18}-\frac{5}9g^3r^{17}M\alpha+g^3 r^{16}M^2\alpha+ \left( {\frac {5}{18}}\alpha g^6+{\frac {4}{27}}g^3 \right) r^{15}\nonumber\\
&  -4/3\,{g}^{3} \left( 4/3+\alpha\,{g}^{3} \right) M{r} ^{14}+ \left( \alpha\,{g}^{3}+8/3 \right) {g}^{3}{M}^{2}{r}^{13}+{ \frac {10}{27}}\,{g}^{6} \left( \alpha\,{g}^{3}+1 \right) {r}^{12}-{ \frac {11}{9}}\, \left( {\frac {30}{11}}+\alpha\,{g}^{3} \right) {g}^{ 6}M{r}^{11}+14/3\,{g}^{6}{r}^{10}{M}^{2}\nonumber\\
& + \left( {\frac {5}{18}}g^ {12}\alpha+{\frac {40}{81}}g^9 \right)r^9-{\frac {7}{18}}\left( \alpha g^3+{\frac {24}{7}} \right) g^9Mr^8+ \left( {\frac {10}{27}} g^{12}+1/9g^{15}\alpha \right) r^6 +2/9g^{12}r^5M+{\frac {1}{54}}g^{15} \left( 8+\alpha g^ 3 \right) r^3+{\frac {2}{81}}g^{18} \bigg] h_1^3\nonumber\\
&+{\frac {1}{81}}h_1 \left( r+g \right) ^6  \left( r^2-gr+{g}^{2} \right) ^6 \bigg]^{-1}{ dr}\bigg\}\left(r^5M -\frac{1}3g^6-\frac{2}3g^3r^3-\frac{1}3r^6 \right) ^{2}\bigg[  \bigg( 3r^{21}\alpha-9M\alpha r^{20}+ \left( 18\alpha g^3+4 \right) r^{18}-90g^3r^{17}M \alpha\nonumber\\
&+162\,{g}^{3}{r}^{16}{M}^{2}\alpha+ \left( 45\,\alpha\,{g}^{6}+ 24\,{g}^{3} \right) {r}^{15}-216\,{g}^{3} \left( 4/3+\alpha\,{g}^{3} \right) M{r}^{14}+162\, \left( \alpha\,{g}^{3}+8/3 \right) g^3M^2r^{13}+ \left( 60\alpha{g}^{9}+60\,{g}^{6} \right) {r}^{12}\nonumber\\
&-9g^6M\left(22\alpha g^3+60\right) r^{11}+756g^6r^{10}M^2+ 5\left( 9g^{12}\alpha+16g^9 \right)r^9-63\left( \alpha g^3+{\frac {24}{7}} \right) g^9Mr^8+ 6g^{12}\left( 10+3g^3\alpha \right) r^6+36g^{12}r^5M\nonumber\\
&+ \left( 24\,{g}^{15}+3\,{g}^{18}\alpha \right) {r}^{3}+4\,{g }^{18} \bigg)  h_1^{2}+27\, \left( r+g \right) ^{3} \left( {r}^{2}-gr+{g}^{2} \right) ^{3} \bigg[ -1/3\,\alpha\,{r}^{12}+M{r}^{11}\alpha- \left( 2/9+\alpha\,{g }^{3} \right) {r}^{9}+M{r}^{8}\alpha\,{g}^{3}\nonumber\\
&+ \left( -2/3\,{g}^{3}- \alpha\,{g}^{6} \right) {r}^{6}+4\,{g}^{3}{r}^{5}M+ \left( -2/3\,{g}^{ 6}-1/3\,\alpha\,{g}^{9} \right) {r}^{3}-2/9\,{g}^{9} \bigg]h_1+2\, \left( r+g \right) ^{6} \left( {r}^{2}-gr+{g}^{ 2} \right) ^{6} \bigg]^{-1}\,.
\end{align}
\begin{align}\label{omega}
&\psi(r)=c_2\,\exp\bigg\{-1/2\,\int \!3\, \bigg[ 1/2\, \bigg\{ \frac{4}9 \left( r+g \right)  \left( {r}^{2}-gr+{g}^{2} \right) r \bigg(M{r}^{11}\alpha -\frac{1}3 \alpha{r}^{12}-\left( \frac{2}9+\alpha g^3 \right) r^9+Mr^8\alpha g^3- \left(\frac{2}3{g}^{3}+\alpha g^6 \right) {r}^{6}\nonumber\\
&+4g^3r^{5}M- \left( 2/3{g}^{6}+1/3\alpha {g}^{9} \right) {r}^{3}-2/9{g}^{9} \bigg)  \left({r}^{5}M-1/3{g}^{6}-2 /3{g}^{3}{r}^{3}-1/3{r}^{6} \right) \chi'' + \chi' \bigg[ \frac{1}9r^{24}\alpha^2-2/3r^{23}M \alpha^2+r^{22}M^2\alpha^2\nonumber\\
&+ \left( 2/3\,{g}^{3}{ \alpha}^{2}+{\frac {20}{243}}\,\alpha \right) {r}^{21}-8/3\, \left( \alpha\,{g}^{3}+1/27 \right) \alpha\,M{r}^{20}+2\,\alpha\, \left( \alpha\,{g}^{3}-2/9 \right) {M}^{2}{r}^{19}+ \left( {\frac {40}{81}}\, \alpha\,{g}^{3}+5/3\,{g}^{6}{\alpha}^{2}+{\frac {44}{729}} \right) {r} ^{18}\nonumber\\
&-4\alpha \left( \alpha g^3+{\frac {56}{81}} \right) g^ 3Mr^{17}+\alpha\left( \alpha\ g^3+{\frac {56}{9}} \right)g^3M^2r^{16}+{\frac {20}{9}}\left( g^6\alpha^2+\frac{5}9 \alpha\ g^3+{\frac {22}{135}} \right) g^3r^{15}-\frac{8}3g^6\alpha  \left( {\frac {26}{9}}+{\frac {38}{27}}+g^3\alpha\right) Mr^{14}\nonumber\\
&+{\frac {20}{3}}\, \left( \alpha\,{g}^{3} +{\frac {32}{45}} \right) {g}^{3}{M}^{2}{r}^{13}+5/3\, \left( {\frac { 44}{81}}+{g}^{6}{\alpha}^{2}+{\frac {80}{81}}\,\alpha\,{g}^{3} \right) {g}^{6}{r}^{12}-2/3\, \left( {g}^{6}{\alpha}^{2}+{\frac {104} {9}}+{\frac {304}{27}}\,\alpha\,{g}^{3} \right) {g}^{6}M{r}^{11}+{ \frac {368}{27}}\,{g}^{6}{r}^{10}{M}^{2}\nonumber\\
&+2/3\, \left( {g}^{6}{\alpha}^ {2}+{\frac {50}{27}}\,\alpha\,{g}^{3}+{\frac {440}{243}} \right) {g}^{ 9}{r}^{9}-{\frac {200}{81}}\,{g}^{9} \left( \alpha\,{g}^{3}+{\frac {42 }{25}} \right) M{r}^{8}+1/9\, \left( {g}^{6}{\alpha}^{2}+{\frac {220}{ 27}}+{\frac {40}{9}}\,\alpha\,{g}^{3} \right) {g}^{12}{r}^{6}-{\frac { 16}{81}}\,{g}^{12}{r}^{5}M\nonumber\\
&+ \left( {\frac {88}{243}}\,{g}^{15}+{\frac {20}{243}}\,{g}^{18}\alpha \right) {r}^{3}+{\frac {44}{729}}\,{g}^{18} \bigg]  \bigg\} \left( r+g \right) ^{2} \left( {r}^{2}-gr+{g}^{2} \right) ^{2}h_1^2+ \bigg[ 4/3\, \bigg( { \frac {1}{54}}\,{r}^{21}\alpha-1/18\,M\alpha\,{r}^{20}+ \left( {\frac {2}{81}}+1/9\,\alpha\,{g}^{3} \right) {r}^{18}\nonumber\\
&-5/9\,{g}^{3}{r}^{17}M \alpha+{g}^{3}{r}^{16}{M}^{2}\alpha+ \left( {\frac {5}{18}}\,\alpha\,{ g}^{6}+{\frac {4}{27}}\,{g}^{3} \right) {r}^{15}-4/3\,{g}^{3} \left( 4 /3+\alpha\,{g}^{3} \right) M{r}^{14}+ \left( \alpha\,{g}^{3}+8/3 \right) {g}^{3}{M}^{2}{r}^{13}\nonumber\\
&+{\frac {10}{27}}\,{g}^{6} \left( \alpha\,{g}^{3}+1 \right) {r}^{12}-{\frac {11}{9}}\, \left( {\frac {30 }{11}}+\alpha\,{g}^{3} \right) {g}^{6}M{r}^{11}+14/3\,{g}^{6}{r}^{10}{ M}^{2}+ \left( {\frac {5}{18}}\,{g}^{12}\alpha+{\frac {40}{81}}\,{g}^{ 9} \right) {r}^{9}-{\frac {7}{18}} \left( \alpha\ g^3+{\frac {24 }{7}} \right) {g}^{9}M{r}^{8}\nonumber\\
&+ \left( {\frac {10}{27}}\,{g}^{12}+1/9\, {g}^{15}\alpha \right) r^6+\frac{2}9{g}^{12}r^5M+{\frac {1}{54}}g^{15} \left( 8+\alpha g^3\right) r^3+{\frac {2}{81}}g^{18} \bigg) r \left(r^5M-\frac{1}3g^6-\frac{2}3g^3r^3-\frac{1}3 r^6 \right)\chi''+ \chi'\bigg\{ {\frac {224}{9}}\,\bigg({M}^{3}  \nonumber\\
&-{\frac {5}{216}}\,\alpha\,{g}^{6}-{\frac {1}{72}}\,{g}^{9}{\alpha}^{2}-\frac{1}{27} {g}^{3} \bigg) {g}^{6}{r}^{15}-{\frac {4}{243}}\,{g}^{24}-{\frac {14}{81}}\, \left( {g}^ {6}{\alpha}^{2}+{\frac {20}{3}}+8/3\,\alpha\,{g}^{3} \right) {g}^{12}{ r}^{12}-{\frac {4}{81}}\, \left( {g}^{6}{\alpha}^{2}+14/3\,\alpha\,{g} ^{3}+{\frac {56}{3}} \right) {g}^{15}{r}^{9}\nonumber\\
&-{\frac {1}{162}}\, \left( {g}^{6}{\alpha}^{2}+{\frac {32}{3}}\,\alpha\,{g}^{3}+{\frac { 224}{3}} \right) {g}^{18}{r}^{6}-{\frac {2}{243}}\,{g}^{21} \left( 16+ \alpha\,{g}^{3} \right) {r}^{3}-1/18\,{r}^{28}{M}^{2}{\alpha}^{2}+{ \frac {248}{27}}\,{g}^{12}{r}^{10}{M}^{2}+1/27\,{r}^{29}M{\alpha}^{2}\nonumber\\
&+ \left(4 \alpha{M}^{3} -{\frac {28}{81}}{g}^{6}{\alpha}^{2}+ \left({M}^{3}{\alpha}^{2} -{\frac {56}{ 243}}\alpha \right) {g}^{3}-{\frac {32}{243}} \right) {g}^{3}{r}^{21}+8\left( -{\frac {35}{648} }{g}^{9}{\alpha}^{2}-{\frac {14}{243}}\alpha{g}^{6}+ \left(\alpha{M}^{3} -{ \frac {14}{243}} \right) {g}^{3}+{\frac {10}{9}}{M}^ {3} \right) {g}^{3}{r}^{18}\nonumber\\
&+{\frac {32}{81}}\,{g}^{18}{r}^{5}M-{\frac {1}{162}}\,{r}^{30}{\alpha}^{2}+ \left( -{\frac {4}{243}}-{\frac {14}{ 81}}\,{g}^{6}{\alpha}^{2}+ \left( {M}^{3}{\alpha}^{2}-{\frac {16}{243} }\,\alpha \right) {g}^{3} \right) {r}^{24}+ \left( -{\frac {4}{81}}\,{ g}^{3}{\alpha}^{2}-{\frac {2}{243}}\,\alpha \right) {r}^{27}+1/3\,{g}^ {3}{r}^{26}M{\alpha}^{2}\nonumber\\
&-{\frac {13}{18}} \left( {\frac {464}{13}}+{ \frac {32}{3}}\alpha g^3+g^6\alpha^2 \right) g^6M^2r^{16}+{\frac {7}{9}}\left( {\frac {320}{63}}+g^6\alpha ^2+{\frac {64}{21}}\alpha g^3 \right) g^9Mr^{14}-{ \frac {46}{27}}\left( {\frac {96}{23}}+\alpha g^3 \right)g^9M^2r^{13}-{\frac {8}{27}} \bigg( \alpha g^3+\frac{7}3 \bigg) \nonumber\\
&\times g^{15}Mr^8+{\frac {4}{27}}g^{12}M \left( g^6\alpha^2-{\frac {28}{3}}-\alpha{g}^{3} \right) {r}^{11}-{\frac {20}{9}}{g}^{3} \left( {g}^{6}\alpha^2+{\frac {64}{15}}+{\frac {23}{5}} \alpha g^3 \right) M^2r^{19}+\frac{5}3 \left( {\frac {136}{45}} \alpha{g}^{3}+{g}^{6}{\alpha}^{2}+{\frac {592}{135}} \right) {g}^{ 6}M{r}^{17}\nonumber\\
&-{\frac {8}{9}}\, \left( \alpha\,{g}^{3}-1/12 \right) \alpha\,{M}^{2}{r}^{25}+{\frac {50}{27}}\,{g}^{3} \left( {\frac {124}{ 75}}+{g}^{6}{\alpha}^{2}+{\frac {52}{25}}\,\alpha\,{g}^{3} \right) M{r }^{20}+{\frac {10}{9}}\,\alpha\,{g}^{3}M \left( \alpha\,{g}^{3}+{ \frac {14}{15}} \right) {r}^{23}-7/3\, \left( \alpha\,{g}^{3}+{\frac { 16}{9}} \right)\nonumber\\
& \alpha\,{g}^{3}{M}^{2}{r}^{22} \bigg\}  \bigg] h_1^3+{\frac {8}{81}}\, \left( r+g \right) ^{5 } \left( {r}^{2}-gr+{g}^{2} \right) ^{5} \bigg\{  \bigg[  \left( -1/18 \,{g}^{9}r-1/6\,{r}^{4}{g}^{6}-1/6\,{r}^{7}{g}^{3}+1/6\,{g}^{3}{r}^{6} M-1/18\,{r}^{10}+1/6\,{r}^{9}M \right)\chi''\nonumber
\end{align}
\begin{align}
&+ \chi' \bigg\{M{r}^{11}\alpha-1/3\,\alpha\,{r}^{12}+ \left( -\alpha\,{g}^{3}-1/6 \right) {r}^{9}+M{r}^{8}\alpha\,{g}^{3}+ \left( -\alpha\,{g}^{6}-1/2\,{g}^{3} \right) {r}^{6}+3\,{g}^{3}{r}^{5 }M+ \left( -1/3\,\alpha\,{g}^{9}-1/2\,{g}^{6} \right) {r}^{3}\nonumber\\
&-\frac{1}{6}{g} ^{9} \bigg\} \bigg] h_1+1/36\, \chi'  \left( r+g \right) ^{3} \left( {r}^{2}-gr+{g}^{2} \right) ^{3} \bigg\}  \bigg] h_1^{-1}\left( -1/3\,{g}^{6}-2/3\,{g}^{3}{r}^{3}+ {r}^{5}M-1/3\,{r}^{6} \right) ^{-1}{r}^{-1} \bigg[  \bigg( {\frac {1}{ 54}}{r}^{21}\alpha-\frac{1}{18}M\alpha\,{r}^{20}\nonumber\\
&+ \left( {\frac {2}{81}}+\frac{1}9\alpha\,{g}^{3} \right) {r}^{18}-\frac{5}9\,{g}^{3}{r}^{17}M\alpha+{g}^{ 3}{r}^{16}{M}^{2}\alpha+ \left( {\frac {5}{18}}\alpha{g}^{6}+{ \frac {4}{27}}\,{g}^{3} \right) {r}^{15}-\frac{4}3\,{g}^{3} \left( \frac{4}3+ \alpha\,{g}^{3} \right) M{r}^{14}+ \left( \alpha\,{g}^{3}+\frac{8}3 \right) {g}^{3}{M}^{2}{r}^{13}\nonumber\\
&+{\frac {10}{27}}\,{g}^{6} \left( \alpha\,{g}^{3 }+1 \right) {r}^{12}-{\frac {11}{9}}\, \left( {\frac {30}{11}}+\alpha \,{g}^{3} \right) {g}^{6}M{r}^{11}+14/3\,{g}^{6}{r}^{10}{M}^{2}+ \left( {\frac {5}{18}}\,{g}^{12}\alpha+{\frac {40}{81}}\,{g}^{9} \right) {r}^{9}-{\frac {7}{18}}\, \left( \alpha\,{g}^{3}+{\frac {24}{ 7}} \right) {g}^{9}M{r}^{8}\nonumber\\
&+ \bigg[ {\frac {10}{27}}\,{g}^{12}+1/9\,{g }^{15}\alpha \bigg] {r}^{6}+2/9\,{g}^{12}{r}^{5}M+{\frac {1}{54}}\,{g }^{15} \left( 8+\alpha\,{g}^{3} \right) {r}^{3}+{\frac {2}{81}}\,{g}^{ 18} \bigg) h_1^2+1/6\, \bigg[  \bigg(M{r}^{11}\alpha-1/3 \alpha{r}^{12}- \left(2/9+\alpha{g}^{3} \right) {r}^{9}\nonumber\\
&+M{r}^{8}\alpha\,{g}^{3}+ \left( -2/3\,{g}^{3}-\alpha\,{g}^{6} \right) {r}^{6}+4\,{g}^{3}{r}^{5}M+ \left( -2/3\,{g}^{6}-1/3\,\alpha \,{g}^{9} \right) {r}^{3}-2/9\,{g}^{9} \bigg) h_1+{\frac {2}{27}}\, \left( r+g \right) ^{3} \left( {r}^{2}-gr+{g}^{ 2} \right) ^{3} \bigg] \nonumber\\
& \left( {r}^{2}-gr+{g}^{2} \right) ^{3} \left( r+g \right) ^{3} \bigg]^{-1} \chi'^{-1}{dr}\bigg\}\,.
\end{align}
Using Eq.~(\ref{sol2}) and (\ref{omega}) in Eq.~({\color{blue}4}), presented in the manuscript, we get:
\begin{align}\label{phi}
&3/2\,\exp\bigg\{-1/2\,\int 3\bigg[ 1/2\, \bigg[ 4/9\, \left( r+g \right)  \left( {r}^{2}-gr+{g}^{2} \right) r \bigg(M{r}^{11}\alpha -\frac{1}3\,\alpha\,{r} ^{12}- \left(\frac{2}9+\alpha\,{g}^{3} \right) {r}^{9}+M{r }^{8}\alpha\,{g}^{3}-\left( \frac{2}3{g}^{3}+\alpha{g}^{6} \right) {r} ^{6}\nonumber\\
&+4{g}^{3}{r}^{5}M- \left( \frac{2}3\,{g}^{6}+\frac{1}3\,\alpha\,{g}^{9} \right) {r}^{3}-2/9\,{g}^{9} \bigg)  \left({r}^{5}M -1/3\,{g}^{6}-2/3\,{g}^{ 3}{r}^{3}-1/3\,{r}^{6} \right) \chi'' + \chi' \bigg\{ \frac{1}9{r}^{24}{\alpha}^{2}-2\frac{2}3{r}^{23}M{ \alpha}^{2}+{r}^{22}{M}^{2}{\alpha}^{2}\nonumber\\
&+ \left( 2/3\,{g}^{3}{\alpha}^{ 2}+{\frac {20}{243}}\,\alpha \right) {r}^{21}-8/3\, \left( \alpha\,{g} ^{3}+1/27 \right) \alpha\,M{r}^{20}+2\,\alpha\, \left( \alpha\,{g}^{3} -2/9 \right) {M}^{2}{r}^{19}+ \left( {\frac {40}{81}}\,\alpha\,{g}^{3} +5/3\,{g}^{6}{\alpha}^{2}+{\frac {44}{729}} \right) {r}^{18}\nonumber\\
&-4\alpha  \left( \alpha{g}^{3}+{\frac {56}{81}} \right) {g}^{3}M{r}^{17}+ \alpha \left( \alpha{g}^{3}+{\frac {56}{9}} \right) {g}^{3}{M}^{2} {r}^{16}+{\frac {20}{9}} \left( {g}^{6}{\alpha}^{2}+\frac{5}9\alpha{g} ^{3}+{\frac {22}{135}} \right) {g}^{3}{r}^{15}-\frac{8}3{g}^{3} \left( { \frac {26}{9}}\alpha{g}^{3}+{\frac {38}{27}}+{g}^{6}{\alpha}^{2} \right) M{r}^{14}\nonumber\\
&+{\frac {20}{3}}\, \left( \alpha\,{g}^{3}+{\frac {32 }{45}} \right) {g}^{3}{M}^{2}{r}^{13}+5/3\, \left( {\frac {44}{81}}+{g }^{6}{\alpha}^{2}+{\frac {80}{81}}\,\alpha\,{g}^{3} \right) {g}^{6}{r} ^{12}-2/3\, \left( {g}^{6}{\alpha}^{2}+{\frac {104}{9}}+{\frac {304}{ 27}}\,\alpha\,{g}^{3} \right) {g}^{6}M{r}^{11}+{\frac {368}{27}}\,{g}^ {6}{r}^{10}{M}^{2}\nonumber\\
&+2/3\, \left( {g}^{6}{\alpha}^{2}+{\frac {50}{27}}\, \alpha\,{g}^{3}+{\frac {440}{243}} \right) {g}^{9}{r}^{9}-{\frac {200} {81}}\,{g}^{9} \left( \alpha\,{g}^{3}+{\frac {42}{25}} \right) M{r}^{8 }+1/9\, \left( {g}^{6}{\alpha}^{2}+{\frac {220}{27}}+{\frac {40}{9}}\, \alpha\,{g}^{3} \right) {g}^{12}{r}^{6}-{\frac {16}{81}}\,{g}^{12}{r}^ {5}M\nonumber\\
&+ \left( {\frac {88}{243}}\,{g}^{15}+{\frac {20}{243}}\,{g}^{18} \alpha \right) {r}^{3}+{\frac {44}{729}}\,{g}^{18} \bigg\}  \bigg] \left( r+g \right) ^{2} \left( {r}^{2}-gr+{g}^{2} \right) ^{2}h_1^2+ \bigg[ 4/3\, \bigg( {\frac {1}{54}}\,{r }^{21}\alpha-1/18\,M\alpha\,{r}^{20}+ \left( {\frac {2}{81}}+1/9\, \alpha\,{g}^{3} \right) {r}^{18}\nonumber\\
&-5/9\,{g}^{3}{r}^{17}M\alpha+{g}^{3}{r }^{16}{M}^{2}\alpha+ \left( {\frac {5}{18}}\,\alpha\,{g}^{6}+{\frac {4 }{27}}\,{g}^{3} \right) {r}^{15}-4/3\,{g}^{3} \left( 4/3+\alpha\,{g}^{ 3} \right) M{r}^{14}+ \left( \alpha\,{g}^{3}+8/3 \right) {g}^{3}{M}^{2 }{r}^{13}\nonumber\\
&+{\frac {10}{27}}\,{g}^{6} \left( \alpha\,{g}^{3}+1 \right) { r}^{12}-{\frac {11}{9}}\, \left( {\frac {30}{11}}+\alpha\,{g}^{3} \right) {g}^{6}M{r}^{11}+14/3\,{g}^{6}{r}^{10}{M}^{2}+ \left( {\frac {5}{18}}\,{g}^{12}\alpha+{\frac {40}{81}}\,{g}^{9} \right) {r}^{9}-{ \frac {7}{18}}\, \left( \alpha\,{g}^{3}+{\frac {24}{7}} \right) {g}^{9 }M{r}^{8}\nonumber\\
&+ \left( {\frac {10}{27}}\,{g}^{12}+1/9\,{g}^{15}\alpha \right) {r}^{6}+2/9\,{g}^{12}{r}^{5}M+{\frac {1}{54}}\,{g}^{15} \left( 8+\alpha\,{g}^{3} \right) {r}^{3}+{\frac {2}{81}}\,{g}^{18} \bigg) r \left( -1/3\,{g}^{6}-2/3\,{g}^{3}{r}^{3}+{r}^{5}M-1/3\,{r}^ {6} \right) \chi''\nonumber\\
& + \chi' \bigg( -{ \frac {4}{243}}\,{g}^{24}+{\frac {224}{9}}\, \left( -{\frac {5}{216}} \,\alpha\,{g}^{6}-{\frac {1}{72}}\,{g}^{9}{\alpha}^{2}-1/27\,{g}^{3}+{ M}^{3} \right) {g}^{6}{r}^{15}-{\frac {14}{81}}\, \left( {g}^{6}{ \alpha}^{2}+{\frac {20}{3}}+8/3\,\alpha\,{g}^{3} \right) {g}^{12}{r}^{ 12}\nonumber\\
&-{\frac {4}{81}}\, \left( {g}^{6}{\alpha}^{2}+14/3\,\alpha\,{g}^{3} +{\frac {56}{3}} \right) {g}^{15}{r}^{9}-{\frac {1}{162}}\, \left( {g} ^{6}{\alpha}^{2}+{\frac {32}{3}}\,\alpha\,{g}^{3}+{\frac {224}{3}} \right) {g}^{18}{r}^{6}-{\frac {2}{243}}\,{g}^{21} \left( 16+\alpha\, {g}^{3} \right) {r}^{3}-1/18\,{r}^{28}{M}^{2}{\alpha}^{2}\nonumber\\
&+{\frac {248} {27}}{g}^{12}{r}^{10}{M}^{2}+\frac{1}{27}{r}^{29}M{\alpha}^{2}+ \left(4\alpha \,{M}^{3} -{ \frac {28}{81}}{g}^{6}{\alpha}^{2}+ \left({M}^{3}{\alpha}^{2} -{\frac {56}{243}} \alpha \right) {g}^{3}-{\frac {32}{243}} \right) {g}^{3}{r}^{21}+8\bigg( {\frac {10}{9}}{M}^{3}-{\frac {35}{648}}{g}^{ 9}{\alpha}^{2}-{\frac {14}{243}}\alpha{g}^{6} \nonumber\\
&+ \left(\alpha{M}^{3}-{\frac {14} {243}} \right) {g}^{3} \bigg) {g}^{3}{r}^{18}+{\frac {32}{81}}{g}^{18}{r}^{5}M-{\frac {1} {162}}{r}^{30}{\alpha}^{2}+ \left[  \left( {M}^{3}{\alpha}^{2}-{\frac {16}{243}} \alpha \right) {g}^{3} -{\frac {4}{243}}-{\frac {14}{81} }{g}^{6}{\alpha}^{2}\right] {r}^{24}-{\frac {2}{243}}\alpha \left(6{g}^ {3}{\alpha}+1 \right) {r}^{27}\nonumber
\end{align}
\begin{align}
&+1/3\,{g}^{3} {r}^{26}M{\alpha}^{2}-{\frac {13}{18}}\, \left( {\frac {464}{13}}+{ \frac {32}{3}}\alpha\,{g}^{3}+{g}^{6}{\alpha}^{2} \right) {g}^{6}{M} ^{2}{r}^{16}+{\frac {7}{9}}\, \left( {\frac {320}{63}}+{g}^{6}{\alpha} ^{2}+{\frac {64}{21}}\alpha{g}^{3} \right) {g}^{9}M{r}^{14}-{ \frac {46}{27}} \left( {\frac {96}{23}}+\alpha{g}^{3} \right) {g}^ {9}{M}^{2}{r}^{13}\nonumber\\
&+{\frac {4}{27}}\,{g}^{12}M \left( -{\frac {28}{3}}+ {g}^{6}{\alpha}^{2}-\alpha\,{g}^{3} \right) {r}^{11}-{\frac {8}{27}}\, \left( \alpha\,{g}^{3}+7/3 \right) {g}^{15}M{r}^{8}-{\frac {20}{9}}\, {g}^{3} \left( {g}^{6}{\alpha}^{2}+{\frac {64}{15}}+{\frac {23}{5}}\, \alpha\,{g}^{3} \right) {M}^{2}{r}^{19}\nonumber\\
&+5/3\, \left( {\frac {136}{45}} \,\alpha\,{g}^{3}+{g}^{6}{\alpha}^{2}+{\frac {592}{135}} \right) {g}^{ 6}M{r}^{17}-{\frac {8}{9}}\, \left( \alpha\,{g}^{3}-1/12 \right) \alpha\,{M}^{2}{r}^{25}+{\frac {50}{27}}\,{g}^{3} \left( {\frac {124}{ 75}}+{g}^{6}{\alpha}^{2}+{\frac {52}{25}}\,\alpha\,{g}^{3} \right) M{r }^{20}\nonumber\\
&+{\frac {10}{9}}\,\alpha\,{g}^{3}M \left( \alpha\,{g}^{3}+{ \frac {14}{15}} \right) {r}^{23}-7/3\, \left( \alpha\,{g}^{3}+{\frac { 16}{9}} \right) \alpha\,{g}^{3}{M}^{2}{r}^{22} \bigg)  \bigg] h_1^3+{\frac {8}{81}}\, \left( r+g \right) ^{5 } \left( {r}^{2}-gr+{g}^{2} \right) ^{5}\nonumber\\
&\times \bigg\{  \bigg[  \left( 1/6\,{g}^{3}{r}^{6} M-1/18 \,{g}^{9}r-1/6\,{r}^{4}{g}^{6}-1/6\,{r}^{7}{g}^{3}-1/18\,{r}^{10}+1/6\,{r}^{9}M \right) \chi'' + \chi' \bigg( -1/3\,\alpha\,{r}^{12}+M{r}^{11}\alpha- \left( \alpha\,{g}^{3}+1/6 \right) {r}^{9}\nonumber\\
&+M{r}^{8}\alpha\,{g}^{3}- \left(\alpha\,{g}^{6}+1/2{g}^{3} \right) {r}^{6}+3{g}^{3}{r}^{5 }M- \left(1/3\alpha{g}^{9}+1/2{g}^{6} \right) {r}^{3}-1/6\,{g} ^{9} \bigg)  \bigg] h_1+1/36\,\chi' \left( r+g \right) ^{3} \left( {r}^{2}-gr+{g}^{2} \right) ^{3} \bigg\} \bigg] h_1^{-1}\bigg\}\nonumber\\
&\bigg[  \bigg( {\frac {1}{54}}\,{r}^{21} \alpha-1/18\,M\alpha\,{r}^{20}+ \left( {\frac {2}{81}}+1/9\,\alpha\,{g }^{3} \right) {r}^{18}-5/9\,{g}^{3}{r}^{17}M\alpha+{g}^{3}{r}^{16}{M}^ {2}\alpha+ \left( {\frac {5}{18}}\,\alpha\,{g}^{6}+{\frac {4}{27}}\,{g }^{3} \right) {r}^{15}\nonumber\\
&-4/3\,{g}^{3} \left( 4/3+\alpha\,{g}^{3} \right) M{r}^{14}+ \left( \alpha\,{g}^{3}+8/3 \right) {g}^{3}{M}^{2}{ r}^{13}+{\frac {10}{27}}\,{g}^{6} \left( \alpha\,{g}^{3}+1 \right) {r} ^{12}-{\frac {11}{9}}\, \left( {\frac {30}{11}}+\alpha\,{g}^{3} \right) {g}^{6}M{r}^{11}+14/3\,{g}^{6}{r}^{10}{M}^{2}\nonumber\\
&+ \left( {\frac {5}{18}}\,{g}^{12}\alpha+{\frac {40}{81}}{g}^{9} \right) {r}^{9}-{ \frac {7}{18}} \left( \alpha\,{g}^{3}+{\frac {24}{7}} \right) {g}^{9 }M{r}^{8}+ \left( {\frac {10}{27}}\,{g}^{12}+\frac{1}9{g}^{15}\alpha \right) {r}^{6}+\frac{2}9{g}^{12}{r}^{5}M+{\frac {1}{54}}{g}^{15} \left( 8+\alpha{g}^{3} \right) {r}^{3}+{\frac {2}{81}}\,{g}^{18} \bigg) h_1^2\nonumber\\
&+1/6\, \bigg(  \bigg[ -1/3\, \alpha\,{r}^{12}+M{r}^{11}\alpha+ \left( -2/9-\alpha\,{g}^{3} \right) {r}^{9}+M{r}^{8}\alpha\,{g}^{3}+ \left( -2/3\,{g}^{3}-\alpha\,{g}^{6} \right) {r}^{6}+4\,{g}^{3}{r}^{5}M+ \left( -2/3\,{g}^{6}-1/3\,\alpha \,{g}^{9} \right) {r}^{3}\nonumber\\
&-\frac{2}9{g}^{9} \bigg] h_1+{\frac {2}{27}}\, \left( r+g \right) ^{3} \left( {r}^{2}-gr+{g}^{ 2} \right) ^{3} \bigg)  \left( {r}^{2}-gr+{g}^{2} \right) ^{3} \left( r+g \right) ^{3} \bigg] ^{-1} \left( -1/3\,{g}^{6}-\frac{2}3{g}^{ 3}{r}^{3}+{r}^{5}M-1/3\,{r}^{6} \right) ^{-1}{r}^{-1} \chi'^{-1}{dr}\bigg\}c_1\nonumber\\
& \bigg\{ \frac{1}2 \bigg[4/9\, \left(\chi' -1 \right)  \left(M{r}^{2}-\frac{1}2\,{g}^{3}-\frac{1}2\,{r}^{3} \right)  \left({r}^{5}M -\frac{1}3{g}^{6}-\frac{2}3{g}^{3}{r}^{3}-\frac{1}3{r}^{ 6} \right) r \left( {r}^{2}-gr+{g}^{2} \right)  \left( r+g \right) \bigg(M{r}^{11}\alpha -\frac{1}3\alpha{r}^{12}+M{r}^{8}\alpha{g}^{3}\nonumber\\
&- \left( \frac{2}9+\alpha{g }^{3} \right) {r}^{9} - \left(\frac{2}3\,{g}^{3}+\alpha{g}^{6} \right) {r}^{6}+4{g}^{3}{r}^{5}M- \left( \frac{2}3\,{g}^{ 6}\frac{1}3\,\alpha{g}^{9} \right) {r}^{3}-\frac{2}9\,{g}^{9} \bigg) \chi''+\chi' \bigg[ {\frac {20}{729}}{g}^{21 }-{\frac {28}{9}}\, \left(\alpha{g}^{3}  -{\frac {25}{378}}\right) \alpha\,M{r}^{23}\nonumber\\
&+{\frac {50}{9}}\, \left( \alpha\,{g}^{3}-{\frac {46} {225}} \right) \alpha\,{M}^{2}{r}^{22}-{\frac {200}{27}}\, \left( {g}^ {6}{\alpha}^{2}+{\frac {94}{225}}+{\frac {53}{75}}\,\alpha\,{g}^{3} \right) {g}^{3}M{r}^{17}+{\frac {38}{9}}\,{g}^{3} \left( {\frac {124} {57}}+{\frac {220}{57}}\,\alpha\,{g}^{3}+{g}^{6}{\alpha}^{2} \right) { M}^{2}{r}^{16}\nonumber\\
&-{\frac {40}{9}}\,{g}^{6} \left( {g}^{6}{\alpha}^{2}+{ \frac {527}{270}}\,\alpha\,{g}^{3}+{\frac {271}{135}} \right) M{r}^{14 }+{\frac {8}{9}}\,{g}^{6} \left( {g}^{6}{\alpha}^{2}+{\frac {259}{18}} \,\alpha\,{g}^{3}+{\frac {85}{3}} \right) {M}^{2}{r}^{13}-4/3\, \left( {\frac {79}{18}}\,\alpha\,{g}^{3}+{g}^{6}{\alpha}^{2}+{\frac { 1472}{243}} \right) {g}^{9}M{r}^{11}\nonumber\\
&+{\frac {220g^9 }{81}}\left( \alpha{g}^{3}+{\frac {274}{55}} \right) {M}^{2}{r}^{10}-{ \frac {4}{27}} \left( {g}^{6}{\alpha}^{2}+{\frac {53}{6}}\alpha\,{ g}^{3}+{\frac {110}{9}} \right) {g}^{12}M{r}^{8}+{\frac {20}{243}}{g }^{15}M \left( \alpha{g}^{3}+{\frac {28}{5}} \right) {r}^{5}+{\frac {14}{27}}\bigg( \frac{\alpha{g}^{3}}3+{\frac {50}{27}}\nonumber\\
&+{g}^{6}{ \alpha}^{2} \bigg) {g}^{12}{r}^{9} \bigg] +{\frac {14}{9}}\,{r}^{25}{M}^{2}{ \alpha}^{2}+{\frac {22}{3}}\, \left( -{\frac {4}{297}}+{\frac {206}{ 297}}\,\alpha\,{g}^{3}+{g}^{6}{\alpha}^{2} \right) {M}^{2}{r}^{19}-{ \frac {16}{27}}\,{r}^{26}M{\alpha}^{2}+{\frac {8}{729}}\,{g}^{18}M{r}^ {2}-{\frac {64}{27}}\,{g}^{12}{M}^{2}{r}^{7}\nonumber\\
&+ \left( -1/2\,{g}^{3}-1/2 \,{r}^{3}+M{r}^{2} \right)  \bigg[ 1/9\,{r}^{24}{\alpha}^{2}-2/3\,{r}^ {23}M{\alpha}^{2}+{r}^{22}{M}^{2}{\alpha}^{2}+ \left( 2/3\,{g}^{3}{ \alpha}^{2}+{\frac {20}{243}}\,\alpha \right) {r}^{21}-8/3\, \left( \alpha\,{g}^{3}+1/27 \right) \alpha\,M{r}^{20}\nonumber\\
&+2\,\alpha\, \left( \alpha\,{g}^{3}-2/9 \right) {M}^{2}{r}^{19}+ \left( {\frac {40}{81}}\, \alpha\,{g}^{3}+5/3\,{g}^{6}{\alpha}^{2}+{\frac {44}{729}} \right) {r} ^{18}-4\,\alpha\, \left( \alpha\,{g}^{3}+{\frac {56}{81}} \right) {g}^ {3}M{r}^{17}+\alpha\, \left( \alpha\,{g}^{3}+{\frac {56}{9}} \right) { g}^{3}{M}^{2}{r}^{16}\nonumber\\
&+{\frac {20}{9}}\, \left( {g}^{6}{\alpha}^{2}+5/9 \,\alpha\,{g}^{3}+{\frac {22}{135}} \right) {g}^{3}{r}^{15}-8/3\,{g}^{ 3} \left( {\frac {26}{9}}\,\alpha\,{g}^{3}+{\frac {38}{27}}+{g}^{6}{ \alpha}^{2} \right) M{r}^{14}+{\frac {20}{3}}\, \left( \alpha\,{g}^{3} +{\frac {32}{45}} \right) {g}^{3}{M}^{2}{r}^{13}\nonumber\\
&+\frac{5}3 \left( {\frac { 44}{81}}+{g}^{6}{\alpha}^{2}+{\frac {80}{81}}\alpha{g}^{3} \right) {g}^{6}{r}^{12}-\frac{2}3\left( {g}^{6}{\alpha}^{2}+{\frac {104} {9}}+{\frac {304}{27}}\alpha{g}^{3} \right) {g}^{6}M{r}^{11}+{ \frac {368}{27}}{g}^{6}{r}^{10}{M}^{2}+\frac{2}3 \left( {g}^{6}{\alpha}^ {2}+{\frac {50}{27}}\alpha\,{g}^{3}+{\frac {440}{243}} \right) {g}^{ 9}{r}^{9}\nonumber\\
&-{\frac {200}{81}}\,{g}^{9} \left( \alpha\,{g}^{3}+{\frac {42 }{25}} \right) M{r}^{8}+1/9\, \left( {g}^{6}{\alpha}^{2}+{\frac {220}{ 27}}+{\frac {40}{9}}\,\alpha\,{g}^{3} \right) {g}^{12}{r}^{6}-{\frac { 16}{81}}\,{g}^{12}{r}^{5}M+ \left( {\frac {88}{243}}\,{g}^{15}+{\frac {20}{243}}\,{g}^{18}\alpha \right) {r}^{3}+{\frac {44}{729}}\,{g}^{18} \bigg]\chi'\nonumber\\
& +{\frac {2}{27}} \left( {\frac {70}{9}}+{\frac {7}{9}}\alpha{g}^{3}+{g}^{6}{\alpha }^{2} \right) {g}^{15}{r}^{6}+{\frac {2}{243}}{g}^{18} \left( \alpha {g}^{3}+{\frac {70}{3}} \right) {r}^{3}-\frac{4}3{g}^{3} \left( -{\frac {35}{18}}{g}^{6}{\alpha}^{2}+ \left( -{\frac {7}{54}}\alpha+{M}^{3 }{\alpha}^{2} \right) {g}^{3}+{\frac {17}{3}}\alpha{M}^{3}-{\frac {35}{243}} \right) {r}^{18}\nonumber\\
&-{\frac {80}{9}}\, \left( -{\frac {7}{24}} \,{g}^{9}{\alpha}^{2}-{\frac {7}{216}}\,\alpha\,{g}^{6}+ \left( -{ \frac {7}{108}}+\alpha\,{M}^{3} \right) {g}^{3}+{\frac {13}{15}}\,{M}^ {3} \right) {g}^{3}{r}^{15}-{\frac {20}{3}}\, \left( {\frac {11}{90}} \,\alpha\,{g}^{3}+{g}^{6}{\alpha}^{2}-{\frac {2}{1215}} \right) M{r}^{ 20}\nonumber\\
&+ \left( {\frac {70}{243}}{g}^{12}\alpha-{\frac {544}{27}}{g}^{ 6}{M}^{3}+{\frac {700}{729}}{g}^{9}+{\frac {14}{9}}{g}^{15}{\alpha }^{2} \right) {r}^{12}+ \left( {\frac {14}{9}}{g}^{6}{\alpha}^{2}+ \left( -\frac{8}3{M}^{3}{\alpha}^{2}+{\frac {14}{243}}\alpha \right) {g }^{3}+\frac{4}3\alpha{M}^{3}+{\frac {20}{729}} \right) {r}^{21}+{\frac { 2}{27}}{r}^{27}{\alpha}^{2}\nonumber
\end{align}
\begin{align}
&+ \left( {\frac {2}{243}}\,\alpha+{\frac {14}{27}}\,{g}^{3}{\alpha}^{2}-4/3\,{M}^{3}{\alpha}^{2} \right) {r}^{ 24} \bigg]  \bigg\} \left( {r}^{2}-gr+{g}^{2} \right) ^{3} \left( r+ g \right) ^{3}h_1^2+ \bigg[ 4/3\, \bigg[ { \frac {1}{54}}\,{r}^{21}\alpha-1/18\,M\alpha\,{r}^{20}+ \left( {\frac {2}{81}}+1/9\,\alpha\,{g}^{3} \right) {r}^{18}\nonumber\\
&-\frac{5}9{g}^{3}{r}^{17}M \alpha+{g}^{3}{r}^{16}{M}^{2}\alpha+ \left( {\frac {5}{18}}\alpha\,{ g}^{6}+{\frac {4}{27}}{g}^{3} \right) {r}^{15}-4/3\,{g}^{3} \left( \frac{4}3+\alpha{g}^{3} \right) M{r}^{14}+ \left( \alpha{g}^{3}+\frac{8}3 \right) {g}^{3}{M}^{2}{r}^{13}+{\frac {10}{27}}{g}^{6} \left( \alpha{g}^{3}+1 \right) {r}^{12}\nonumber\\
&-{\frac {11}{9}}\, \left( {\frac {30 }{11}}+\alpha\,{g}^{3} \right) {g}^{6}M{r}^{11}+14/3\,{g}^{6}{r}^{10}{ M}^{2}+ \left( {\frac {5}{18}}\,{g}^{12}\alpha+{\frac {40}{81}}\,{g}^{ 9} \right) {r}^{9}-{\frac {7}{18}}\, \left( \alpha\,{g}^{3}+{\frac {24 }{7}} \right) {g}^{9}M{r}^{8}+ \left( {\frac {10}{27}}\,{g}^{12}+1/9\, {g}^{15}\alpha \right) {r}^{6}\nonumber\\
&+\frac{2}9\,{g}^{12}{r}^{5}M+{\frac {1}{54}}\, {g}^{15} \left( 8+\alpha\,{g}^{3} \right) {r}^{3}+{\frac {2}{81}}\,{g} ^{18} \bigg]  \left(\chi' -1 \right)  \left( -1/2\,{g}^{3}-1/2\,{r}^{3}+M{r}^{2} \right)  \left( - 1/3\,{g}^{6}-2/3\,{g}^{3}{r}^{3}+{r}^{5}M-1/3\,{r}^{6} \right) r \nonumber\\
&\left( {r}^{2}-gr+{g}^{2} \right)  \left( r+g \right) \chi'' + \chi' \bigg(  \left( {\frac {2}{243}}\,{g}^ {30}\alpha-{\frac {40}{729}}\,{g}^{27} \right) {r}^{3}-{\frac {284}{81 }}\,{g}^{18}M \left( {\frac {398}{213}}+\alpha\,{g}^{3} \right) {r}^{ 11}+{\frac {104}{81}}\, \left( \alpha\,{g}^{3}+{\frac {421}{39}} \right) {g}^{18}{M}^{2}{r}^{10}\nonumber\\
&-{\frac {58}{81}}\,{g}^{21} \left( \alpha\,{g}^{3}+{\frac {20}{29}} \right) M{r}^{8}-{\frac {8}{243}}\,M{ g}^{24} \left( \alpha\,{g}^{3}-12 \right) {r}^{5}-\frac{4}3{g}^{3} \left( \alpha\,{M}^{3}+{\frac {47}{54}}\alpha {g}^{3}-{\frac {44}{27}} \right) M{r}^{26}-{\frac {16}{27}}\,{g}^{3} \left( {\frac {166}{9}}+ \alpha{g}^{3} \right) {M}^{2}{r}^{25}\nonumber\\
&-\frac{8}3 \left( {\frac {67}{54}} \alpha{g}^{6}+ \left( \alpha{M}^{3}-{\frac {85}{27}} \right) {g} ^{3}+16/3{M}^{3} \right) {g}^{3}M{r}^{23}+{\frac {232}{81}} \left(\alpha{g}^{3} -{\frac {425}{29}} \right) {g}^{6}{M}^{2}{r}^{ 22}-\frac{4}3{g}^{6} \bigg( {\frac {293}{54}}\alpha{g}^{6}+ \left( \alpha\,{M}^{3}-{\frac {574}{81}} \right) {g}^{3}\nonumber\\
&+{\frac {106}{3}}\,{M }^{3} \bigg) M{r}^{20}+{\frac {868}{81}}{g}^{9} \left( -{\frac { 2612}{651}}+\alpha{g}^{3} \right) {M}^{2}{r}^{19}-{\frac {224}{9}}\, \left( {M}^{3}+{\frac {25}{63}}\alpha{g}^{6}+{\frac {19}{252}}{ g}^{3} \right) {g}^{9}M{r}^{17}+{\frac {344}{27}}{g}^{12} \left( { \frac {365}{387}}+\alpha{g}^{3} \right) {M}^{2}{r}^{16}\nonumber\\
&-{\frac {214} {27}}\, \left( {\frac {436}{321}}+\alpha\,{g}^{3} \right) {g}^{15}M{r} ^{14}+{\frac {536}{81}}\,{g}^{15} \left( {\frac {394}{67}}+\alpha\,{g} ^{3} \right) {M}^{2}{r}^{13}+4/9\, \left( {\frac {35}{9}}\,\alpha\,{g} ^{6}+ \left( \alpha\,{M}^{3}-{\frac {40}{27}} \right) {g}^{3}+{\frac { 518}{3}}\,{M}^{3} \right) {g}^{6}{r}^{21}\nonumber\\
&-16/3\, \left( -{\frac {7}{18 }}\,\alpha\,{g}^{6}+ \left( {\frac {35}{162}}+\alpha\,{M}^{3} \right) {g}^{3}-{\frac {37}{3}}\,{M}^{3} \right) {g}^{9}{r}^{18}+{\frac {20}{ 81}}\,{r}^{31}{M}^{2}\alpha-{\frac {20}{243}}\,{r}^{32}\alpha\,M-{ \frac {256}{243}}\,{M}^{2}{r}^{7}{g}^{21}-{\frac {8}{729}}\,{g}^{27}M{ r}^{2}\nonumber\\
&+4/3 \left( \alpha{M}^{3}+{\frac {5}{18}}\alpha{g}^{3}-{ \frac {10}{243}} \right) {g}^{3}{r}^{27}+{\frac {8}{81}} \left( 1/3+ \alpha{g}^{3} \right) {M}^{2}{r}^{28}+{\frac {20}{243}}{g}^{24} \left( \alpha{g}^{3}-3 \right) {r}^{6}-{\frac {28}{9}}{g}^{12} \bigg[ \left( \alpha{M}^{3}+\frac{4}9 \right) {g}^ {3} -5/9\alpha{g}^{6}\nonumber\\
&+{\frac {106}{21}}{M}^{3} \bigg] {r}^{15}-{\frac {688}{27}} \left( {M}^{3}-{\frac {5}{129}}\alpha{g}^{6}+{\frac {35{g}^{3}}{774}} \right) {g}^{15}{r}^{12}+{\frac {10}{27}}\left(\alpha{g}^{3} -{\frac {16 }{9}}\right) {g}^{21}{r}^{9}-{\frac {4}{729}}{g}^{ 30}+ \left( {\frac {20}{243}} \alpha{g}^{3}-\frac{2}9\alpha{M}^{3}-{\frac {4}{729}} \right) {r}^{30}\nonumber\\
&+ \left( \alpha{M}^{3}+\frac{4}9 \right) {g}^ {3}+ \left( -{\frac {32}{81}}\,M\alpha\,{ g}^{3}-{\frac {8}{729}}\,M \right) {r}^{29}+{\frac {2}{243}}\,{r}^{33} \alpha+ \left( {\frac {80}{81}}\,\alpha\,{g}^{9}+ \left( -{\frac {20}{ 81}}+{\frac {38}{9}}\,\alpha\,{M}^{3} \right) {g}^{6}+{\frac {560}{27} }\,{g}^{3}{M}^{3} \right) {r}^{24}\nonumber\\
&+ \bigg[ {\frac {224}{9}} \left( {M}^{3}-{\frac {5}{216}}\,\alpha\,{g}^{6}-{\frac { 1}{72}}\,{g}^{9}{\alpha}^{2}-1/27\,{g}^{3} \right) {g}^{6}{r}^ {15}-{\frac {4}{243}}{g}^{24} -{\frac {14}{81}}\, \left( {g}^{6}{\alpha}^{2}+{\frac {20}{3}}+8/3 \,\alpha\,{g}^{3} \right) {g}^{12}{r}^{12}\nonumber\\
&-{\frac {4 g^{ 15}}{81}} \left( g ^6\alpha^2+\frac{14}3\alpha g^3+{\frac {56}{3}} \right)r^9-{\frac {g^{18}}{162}} \left( {g}^{6}{\alpha}^{2}+{\frac {32}{3 }}\alpha{g}^{3}+{\frac {224}{3}} \right) r^6-{\frac {2g^{21} }{243}} \left( 16+\alpha{g}^{3} \right) {r}^{3}-\frac{1}{18}{r}^ {28}{M}^{2}{\alpha}^{2}+{\frac {248}{27}}{g}^{12}{r}^{10}{M}^{2}\nonumber\\
&+1/ 27\,{r}^{29}M{\alpha}^{2}+ \left( -{\frac {28}{81}}\,{g}^{6}{\alpha}^{ 2}+ \left( -{\frac {56}{243}}\,\alpha+{M}^{3}{\alpha}^{2} \right) {g}^ {3}-{\frac {32}{243}}+4\,\alpha\,{M}^{3} \right) {g}^{3}{r}^{21}+8\, \bigg( -{\frac {35}{648}}\,{g}^{9}{\alpha}^{2}-{\frac {14}{243}}\, \alpha\,{g}^{6}+{\frac {10}{9}}\,{M}^{3} \nonumber\\
&+ \left( -{\frac {14}{243}}+\alpha\,{M}^{3} \right) {g} ^{3}\bigg) {g}^{3}{r}^{18}+{\frac {32}{81}} \,{g}^{18}{r}^{5}M-{\frac {1}{162}}\,{r}^{30}{\alpha}^{2}+ \left( -{ \frac {4}{243}}-{\frac {14}{81}}\,{g}^{6}{\alpha}^{2}+ \left( {M}^{3}{ \alpha}^{2}-{\frac {16}{243}}\,\alpha \right) {g}^{3} \right) {r}^{24} \nonumber\\
&- \left( {\frac {4}{81}}{g}^{3}{\alpha}^{2}+{\frac {2}{243}}\, \alpha \right) {r}^{27}+\frac{1}3{g}^{3}{r}^{26}M{\alpha}^{2}-{\frac {13}{ 18}}\, \left( {\frac {464}{13}}+{\frac {32}{3}}\,\alpha\,{g}^{3}+{g}^{ 6}{\alpha}^{2} \right) {g}^{6}{M}^{2}{r}^{16}+{\frac {7}{9}}\, \left( {\frac {320}{63}}+{g}^{6}{\alpha}^{2}+{\frac {64}{21}}\,\alpha\,{g}^{3 } \right) {g}^{9}M{r}^{14}\nonumber\\
&-{\frac {46}{27}} \left( {\frac {96}{23}}+ \alpha{g}^{3} \right) {g}^{9}{M}^{2}{r}^{13}+{\frac {4M}{27}}{g}^{ 12} \left( {g}^{6}{\alpha}^{2}-{\frac {28}{3}}-\alpha{g}^{3} \right) {r}^{11}-{\frac {20}{9}}g^3 \left( g^6 \alpha^2+{\frac {64}{15}}+{\frac {23}{5}}\alpha{g}^{3} \right) {M}^{2}{r}^{19}+\frac{5}3 \left( {\frac {136}{45}}\alpha\,{g}^{3}\right.\nonumber\\
&\left.+{g}^{6 }{\alpha}^{2}+{\frac {592}{135}} \right) {g}^{6}M{r}^{17}-{\frac {8}{9 }} \left( \alpha{g}^{3}-\frac{1}{12} \right) \alpha{M}^{2}{r}^{25}-{\frac {8M}{27}} \left( \alpha{g}^{3}+\frac{7}3 \right) {g}^{15}{r}^{8}+{ \frac {50}{27}}{g}^{3} \left( {\frac {124}{75}}+{g}^{6}{\alpha}^{2}+ {\frac {52}{25}}\alpha{g}^{3} \right) M{r}^{20}\nonumber\\
&+{\frac {10}{9}}\, \alpha\,{g}^{3}M \left( \alpha\,{g}^{3}+{\frac {14}{15}} \right) {r}^{ 23}-7/3\, \left( \alpha\,{g}^{3}+{\frac {16}{9}} \right) \alpha\,{g}^{ 3}{M}^{2}{r}^{22} \bigg]  \left( -1/2\,{g}^{3}-1/2\,{r}^{3}+M{r}^{2} \right)  \left( {r}^{2}-gr+{g}^{2} \right)  \left( r+g \right) \chi'\bigg)  \bigg] h_1^3\nonumber\\
&+{\frac {8}{81}}\bigg[  \bigg( 1/6\, \left( \chi' -1 \right)  \left( -1/2\,{g} ^{3}-1/2\,{r}^{3}+M{r}^{2} \right)  \left( -1/3\,{g}^{6}-2/3\,{g}^{3}{ r}^{3}+{r}^{5}M-1/3\,{r}^{6} \right) r \left( {r}^{2}-gr+{g}^{2} \right)  \left( r+g \right) \chi''\nonumber\\
&+ \bigg[  \left( -1/2\,{g}^{3}-1/2\,{r}^{3}+M{r}^{2} \right)  \bigg( -1/3\,\alpha\,{r}^{12}+M{r}^{11}\alpha+ \left( - \alpha\,{g}^{3}-1/6 \right) {r}^{9}+M{r}^{8}\alpha\,{g}^{3}+ \left( - \alpha\,{g}^{6}-1/2\,{g}^{3} \right) {r}^{6}+3\,{g}^{3}{r}^{5}M\nonumber\\
&- \left( \frac{1}3\,\alpha\,{g}^{9}+1/2\,{g}^{6} \right) {r}^{3}-\frac{1}6\,{g}^{9 } \bigg)\chi' -{\frac {7}{24}} \,{r}^{15}\alpha+{\frac {35}{24}}\,{r}^{14}M\alpha-\frac{7}4\,{r}^{13}{M}^{2 }\alpha- \left( \frac{7}6\,\alpha\,{g}^{3}+\frac{1}9 \right) {r}^{12}+ \left( \frac{1}9 M+\frac{7}2\,M\alpha{g}^{3} \right) {r}^{11}\nonumber\\
&-\frac{7}4\, \left(\alpha\, {g}^{3} -\frac{2}{21} \right) {M}^{2}{r}^{10}- \left( \frac{4}9\,{g}^{3}+\frac{7}4\,\alpha\,{g} ^{6} \right) {r}^{9}+{\frac {21}{8}}\, \left( \alpha\,{g}^{3}+{\frac { 80}{63}} \right) {g}^{3}M{r}^{8}-16/3\,{g}^{3}{M}^{2}{r}^{7}+- \left( \frac{7}6\,\alpha\,{g}^{9}+\frac{2}3\,{g}^{6} \right) {r}^{6}\nonumber\\
&+{\frac {7}{12}}\,{g} ^{6} \left( {\frac {40}{7}}+\alpha\,{g}^{3} \right) M{r}^{5}+ \left( - {\frac {7}{24}}\,{g}^{12}\alpha-4/9\,{g}^{9} \right) {r}^{3}+1/9\,{g}^ {9}M{r}^{2}-1/9\,{g}^{12} \bigg]\chi'  \bigg) h_1+1/36\, \left( \chi' -2 \right)  \chi'  \left(M{r}^{2} -1/2\,{r}^{3}\right. \nonumber
\end{align}
\begin{align}
& \left. -\frac{1}2\,{g}^{3} \right)\left( {r}^{2}-gr+{g}^{2} \right) ^{3} \left( r+g \right) ^{3} \bigg]  \left( {r}^{2}-gr+{g}^{2} \right) ^{6} \left( r +g \right) ^{6}  \bigg[  \bigg( {\frac {1}{54}}r^{21} \alpha-\frac{1}{18}M\alpha{r}^{20}+ \left( {\frac {2}{81}}+\frac{1}9\,\alpha\,{g }^{3} \right) {r}^{18}-\frac{5}9{g}^{3}{r}^{17}M\alpha\nonumber\\
&+{g}^{3}{r}^{16}{M}^ {2}\alpha+ \left( {\frac {5}{18}}\,\alpha\,{g}^{6}+{\frac {4}{27}}\,{g }^{3} \right) {r}^{15}-4/3\,{g}^{3} \left( 4/3+\alpha\,{g}^{3} \right) M{r}^{14}+ \left( \alpha\,{g}^{3}+8/3 \right) {g}^{3}{M}^{2}{ r}^{13}+{\frac {10}{27}}\,{g}^{6} \left( \alpha\,{g}^{3}+1 \right) {r} ^{12}\nonumber\\
&-{\frac {11}{9}}\, \left( {\frac {30}{11}}+\alpha\,{g}^{3} \right) {g}^{6}M{r}^{11}+14/3\,{g}^{6}{r}^{10}{M}^{2}+ \left( {\frac {5}{18}}\,{g}^{12}\alpha+{\frac {40}{81}}\,{g}^{9} \right) {r}^{9}-{ \frac {7}{18}}\, \left( \alpha\,{g}^{3}+{\frac {24}{7}} \right) {g}^{9 }M{r}^{8}+ \left( {\frac {10}{27}}\,{g}^{12}+1/9\,{g}^{15}\alpha \right) {r}^{6}\nonumber\\
&+2/9\,{g}^{12}{r}^{5}M+{\frac {1}{54}}\,{g}^{15} \left( 8+\alpha\,{g}^{3} \right) {r}^{3}+{\frac {2}{81}}\,{g}^{18} \bigg) h_1^2+1/6\, \bigg[  \bigg( -1/3\, \alpha\,{r}^{12}+M{r}^{11}\alpha+ \left( -2/9-\alpha\,{g}^{3} \right) {r}^{9}+M{r}^{8}\alpha\,{g}^{3}\nonumber\\
&+ \left( -2/3\,{g}^{3}-\alpha\,{g}^{6} \right) {r}^{6}+4\,{g}^{3}{r}^{5}M+ \left( -2/3\,{g}^{6}-1/3\,\alpha \,{g}^{9} \right) {r}^{3}-2/9\,{g}^{9} \bigg) h_1+{\frac {2}{27}}\, \left( r+g \right) ^{3} \left( {r}^{2}-gr+{g}^{ 2} \right) ^{3} \bigg]  \left( {r}^{2}-gr+{g}^{2} \right) ^{3}\nonumber\\
& \left( r+g \right) ^{3} \bigg] ^{-1}  \left({r}^{5}M-\frac{1}3{g}^{6}-\frac{2}3{g}^{ 3}{r}^{3}-\frac{1}3{r}^{6} \right) ^{-1}{r}^{-1}\left( {r}^{2}-g r+{g}^{2} \right) ^{-2} \left( r+g \right) ^{-2}=0\,.
\end{align}
The above equation is very difficult to solve it but its asymptotic solution as $r\to 0$ takes the form:
\begin{align}\label{pa}
\chi=6r+c_3\,.
\end{align}
{\color{blue} The invariants of the solution of the case with unequal metric components take the form:}
\begin{align}\label{K2}
&K=R_{\mu \nu \rho \sigma}R^{\mu \nu \rho \sigma}=\bigg[  \bigg( {\alpha}^{2}\{ 18{r}^{24}-72{r}^{23}M+81{r}^{22}{M}^{2}+108{r}^{21}{g}^{3}\}+36M\alpha \left(1 -15{g}^{3}{\alpha}\right) {r}^{20}+ M^2\left( 270{g}^{3}{\alpha}^{2}-72\alpha \right) {r}^{19}\nonumber\\
&+\left( 4+270{g}^{6}{\alpha}^{2} \right) {r}^{18}+ 16M\left(3\alpha  g^3 -1-15g^6\alpha^2 \right) r^{17}+ 3M^2\left( 135g^6 \alpha^2+16+48g^3 \alpha \right) r^{16}+  24g^3\left(1+15g^6\alpha^2 \right) r^ {15}\nonumber\\
&-16M \bigg(5+3\alpha g^3+45g^6\alpha^2 \bigg) g^3r^{14}+ 4M^2{g}^{3}\left( 81{\alpha}^{2}{g}^{6}-48 \right) {r}^{13}+ 600{g}^{6}\left( 6+27{\alpha}^{2}{g}^{6} \right) {r}^{12}-36 M{g}^{6} \bigg( {\frac {8}{15}}\alpha{g}^{3}+\frac{4}9+{g}^{6}{\alpha}^{2} \bigg) {r}^{11}\nonumber\\
&+108\,{M}^{2} \left( -\frac{2}3\alpha{g}^{3}+8+{g}^{6}{\alpha}^{2} \right) {g}^{6}{r}^{10}+ \left( 80{g}^{9}+108\,{\alpha}^{2}{g}^{15} \right) {r}^{9}-72 \left( {\frac {20}{9}}+\frac{7}3\alpha{g}^{3}+{g}^{6}{\alpha}^ {2} \right) M{g}^{9}{r}^{8}+144{M}^{2} \left( \alpha{g}^{3}-\frac{2}{3} \right) { g}^{9}{r}^{7}\nonumber\\
&+ \left( 60\,{g}^{12}+18\,{\alpha}^{2}{g}^{18} \right) {r}^{6} -48\,M{g}^{12} \left( \alpha{g}^{3}+5/3 \right) {r}^{5}+96\,{r}^{4}{M}^{2}{ g}^{12}+24\,{g}^{15}{r}^{3}-16\,{g}^{15}M{r}^{2}+4\,{g}^{18} \bigg) { e^{2\,\alpha{r}^{3}}}+16\, \left( r+g \right) ^{5}\nonumber\\
& \left( {r}^{2}-rg+ {g}^{2} \right) ^{5} \left(  \left( -1/2\,{g}^{3}-1/2\,{r}^{3}+M{r}^{2 } \right) {e^{\alpha{r}^{3}}}+1/4\,{r}^{3}+1/4\,{g}^{3} \right) \bigg]  \left( {r}^{3}+{g}^{3} \right) ^{-6}{r}^{-4}\,.
\end{align}
The Ricci tensor squared takes the form
\begin{align}\label{rts2}
&R_{\mu \nu }R^{\mu \nu}=\frac{1}{2\left( {r}^{3}+{g}^{3} \right) ^6{r}^4} \bigg[  \bigg( 27\,{r}^{24}{\alpha}^{2}-90\,{r}^{23}{\alpha}^{2}M+81\,{r}^ {22}{M}^{2}{\alpha}^{2}+ \left( 162\,{g}^{3}{\alpha}^{2}+12\,\alpha \right) {r}^{21}+ \left( -504\,{g}^{3}M{\alpha}^{2}-24\,M\alpha \right) {r}^{20}\nonumber\\
&+378\,{g}^{3}{r}^ {19}{M}^{2}{\alpha}^{2}+ \left( 72\,\alpha{g}^{3}+405\,{g}^{6}{\alpha}^{2}+4 \right) {r}^{18}+ \left( -1116\,{g}^{6}M{\alpha}^{2}-48\,\alpha M{g}^{3} \right) {r}^{17} +729\,{g}^{6}{r}^{16}{M}^{2}{\alpha}^{2}+ \bigg( 180\,\alpha{g}^{6}+24\,{g}^{3}\nonumber\\
&+ 540\,{g}^{9}{\alpha}^{2} \bigg) {r}^{15}-1224\,M \left( {g}^{6}{\alpha}^{2}+{ \frac {7}{51}}\,\alpha{g}^{3}+{\frac {2}{51}} \right) {g}^{3}{r}^{14}+648\, \left( \alpha{g}^{3}-1/3 \right) {M}^{2}\alpha{g}^{6}{r}^{13}+ \left( 240\,\alpha{g} ^{9}+60\,{g}^{6}+405\,{\alpha}^{2}{g}^{12} \right) {r}^{12}\nonumber\\
&-666M \left( { \frac {76}{111}}\alpha{g}^{3}+{g}^{6}{\alpha}^{2}+{\frac {32}{111}} \right) {g }^{6}{r}^{11}+216{M}^{2} \left( 10/3+{g}^{6}{\alpha}^{2}+\alpha{g}^{3} \right) {g}^{6}{r}^{10}+ {g}^{9}\left( 80+162{\alpha}^{2}{g}^6+180 {g}^3\alpha \right) {r}^{9}\nonumber\\
&-144\,M{g}^{9} \left( 2+{g}^{6}{\alpha}^{2}+\frac{10}3 \alpha{g}^{3} \right) {r}^{8}+432{M}^{2} \left( \alpha{g}^{3}-\frac{2}3 \right) {g }^{9}{r}^{7}+ \left( 27\,{\alpha}^{2}{g}^{18}+72\,{g}^{15}\alpha+60\,{g}^{12} \right) {r}^{6}-168\, \left( \alpha{g}^{3}+{\frac {8}{7}} \right) M{g}^{12 }{r}^{5}\nonumber\\
&+288\,{M}^{2}{r}^{4}{g}^{12}+ \left( 12\,{g}^{18}\alpha+24\,{g}^{15 } \right) {r}^{3}-48\,{g}^{15}M{r}^{2}+4\,{g}^{18} \bigg) {^{2 \,\alpha{r}^{3}}}+24\, \bigg(  \bigg( -1/2\,{r}^{9}\alpha+M{r}^{8}\alpha+ \left( -\alpha{g }^{3}-1/3 \right) {r}^{6}+{g}^{3}\alpha{r}^{5}M\nonumber\\
&+ \left( -1/2\,\alpha{g}^{6}-2/3 \,{g}^{3} \right) {r}^{3}+2\,{g}^{3}M{r}^{2}-1/3\,{g}^{6} \bigg) { ^{\alpha{r}^{3}}}+1/3\,{g}^{3}{r}^{3}+1/6\,{g}^{6}+1/6\,{r}^{6} \bigg)  \left( {r}^{2}-rg+{g}^{2} \right) ^{4} \left( r+g \right) ^{ 4} \bigg]\,.
\end{align}
Finally, the Ricci scalar yields the form:
\begin{align}\label{ricci2}
&R=\frac {1}{ \left( {r}^{3}+{g}^{3} \right) ^{3}{r}^{2
}} \bigg(9\,M{r}^{11}\alpha -6\,\alpha{r}^{12}- 2\left( 9\alpha{g}^{3}+1
 \right) {r}^{9}+27\,M{r}^{8}\alpha{g}^{3}-6\left( {g}^{3}+3\alpha{g}^{6}
 \right) {r}^{6}+ 6\left( 3M\alpha{g}^{6}-2M{g}^{3} \right) {r}^{5}\nonumber\\
& -6{g}^{6}
 \left( \alpha{g}^{3}+1 \right) {r}^{3}+24\,{g}^{6}M{r}^{2}-2
\,{g}^{9} \bigg) {e^{\alpha{r}^{3}}}+6\,{r}^{6}{g}^{3}+6\,{r}^{3}{g}
^{6}+2\,{r}^{9}+2\,{g}^{9}\,.
\end{align}
\section{The three roots of $g_{tt}$ given by Eq.~(13)}
The roots of Eq.~({\color{blue} 13}), presented in the manuscript, has the form:
\begin{align}\label{roots}
&r_0=\frac{1}{6}\sqrt [3]{-108\,{g}^{3}+64\,{M}^{3}+12\,\sqrt {81\,{g}^{6}-96\,{g}^{3}{M}^{3}}}+\frac{8}3\,{\frac {{M}^{2}}{\sqrt [3]{-108\,{g}^{3}+64\,{M}^{
3}+12\,\sqrt {81\,{g}^{6}-96\,{g}^{3}{M}^{3}}}}}+\frac{2}3\,M\,,\nonumber\\
&r_1=-\frac{1}{12}\,\sqrt [3]{-108\,{g}^{3}+64\,{M}^{3}+12\,\sqrt {81\,{g}^{6}-96\,
{g}^{3}{M}^{3}}}-\frac{4}3\,{\frac {{M}^{2}}{\sqrt [3]{-108\,{g}^{3}+64\,{M}
^{3}+12\,\sqrt {81\,{g}^{6}-96\,{g}^{3}{M}^{3}}}}}+\frac{2}3\,M\nonumber\\
&+\frac{i}2\sqrt
{3} \left( \frac{1}6\,\sqrt [3]{-108\,{g}^{3}+64\,{M}^{3}+12\,\sqrt {81\,{g}
^{6}-96\,{g}^{3}{M}^{3}}}-\frac{8}3\,{\frac {{M}^{2}}{\sqrt [3]{-108\,{g}^{3
}+64\,{M}^{3}+12\,\sqrt {81\,{g}^{6}-96\,{g}^{3}{M}^{3}}}}} \right)\,,\nonumber\\
&r_2=-\frac{1}{12}\,\sqrt [3]{-108\,{g}^{3}+64\,{M}^{3}+12\,\sqrt {81\,{g}^{6}-96\,
{g}^{3}{M}^{3}}}-\frac{4}3\,{\frac {{M}^{2}}{\sqrt [3]{-108\,{g}^{3}+64\,{M}
^{3}+12\,\sqrt {81\,{g}^{6}-96\,{g}^{3}{M}^{3}}}}}+\frac{2}3\,M\nonumber\\
&-\frac{i}2\sqrt
{3} \left( 1/6\,\sqrt [3]{-108\,{g}^{3}+64\,{M}^{3}+12\,\sqrt {81\,{g}
^{6}-96\,{g}^{3}{M}^{3}}}-\frac{8}3\,{\frac {{M}^{2}}{\sqrt [3]{-108\,{g}^{3
}+64\,{M}^{3}+12\,\sqrt {81\,{g}^{6}-96\,{g}^{3}{M}^{3}}}}} \right)
\end{align}
The reality of the above roots depends on with the quantity $81\,{g}^{6}-96\,{g}^{3}{M}^{3}$ greater/less than zero. In the case of this quantity positive then we have one real roots and in case it has a negative value then we have two real roots as shown in Figure~{\color{blue} 1(b)}.